\documentclass[letterpaper,11pt,fleqn]{article}
\usepackage{jheppub}

\usepackage{graphicx}
\usepackage{bm,amsmath,amssymb}
\usepackage[mathscr]{eucal}

\long\def\comment#1{ }
\newcommand{\eqn}[1]{Eq.~(\ref{#1})}
\newcommand{\beq}{\begin{equation}}
\newcommand{\eeq}{\end{equation}}
\newcommand{\nn}{\nonumber\\}
\newcommand{\dif}{{\rm d}}

\newcommand{\rme}{{\rm e}}
\newcommand{\rmi}{{\rm i}}
\newcommand{\rmP}{{\rm P}}

\newcommand{\rmTr}{{\rm Tr}}
\newcommand{\del}{\partial}

\newcommand{\mcal}{\mathcal}

\newcommand{\bk}{\bm{k}}

\newcommand{\bp}{\bm{p}}
\newcommand{\bx}{\bm{x}}
\newcommand{\by}{\bm{y}}
\newcommand{\bu}{\bm{u}}
\newcommand{\bv}{\bm{v}}
\newcommand{\bz}{\bm{z}}
\newcommand{\bw}{\bm{w}}

\newcommand{\bbx}{\bm{\bar{x}}}
\newcommand{\bby}{\bm{\bar{y}}}

\newcommand{\bbz}{\bm{\bar{z}}}

\newcommand{\abar}{\bar{\alpha}}
\newcommand{\pd}{{\phantom{\dagger}}}

\title{\Large JIMWLK evolution for multi-particle production in Langevin form}

\author[a]{E.~Iancu}
\author[b]{and D.N.~Triantafyllopoulos}

\affiliation[a]{Institut de Physique Th\'{e}orique de Saclay,
F-91191 Gif-sur-Yvette, France}
\affiliation[b]{European Centre for Theoretical Studies in Nuclear Physics and Related Areas (ECT*)\\ and Fondazione Bruno Kessler, Strada delle Tabarelle 286, I-38123 Villazzano (TN), Italy}

\emailAdd{edmond.iancu@cea.fr}
\emailAdd{trianta@ectstar.eu}

\abstract{Within the effective theory for the Color Glass Condensate, we study multi-particle production
with rapidity correlations in proton-nucleus collisions at high energy. The high-energy evolution 
responsible for such correlations is governed by a generalization of the JIMWLK equation
which describes the simultaneous evolution of the (strong) nuclear color fields in the direct amplitude
and the complex conjugate amplitude. This functional equation can be used to derive
ordinary evolution equations for the cross-sections for particle production (a generalization of the
Balitsky hierarchy). However, the ensuing equations appear to
be too complicated to be useful in practice, including in the limit where the number of colors is large.
To circumvent this problem, we propose an alternative formulation of the high-energy
evolution as a Langevin process, which is better suited for numerical implementations. 
This process is directly oriented towards the calculation of the cross-sections,
so its detailed structure depends upon the nature of the final state.
We present the stochastic equations appropriate for two gluon production,
and also for three gluon production, with generic rapidity differences.
}

\begin{document}
\maketitle

\section{Introduction}
\label{sect-intro}

The study of multi-particle production in hadron-hadron collisions at very high energy,
as currently pursued at RHIC and the LHC, provides strong evidence in favour of collective
phenomena associated with high parton densities. In the case of proton-proton or proton-nucleus
collisions, such phenomena are generally attributed to `initial-state interactions', that is,
to the existence of high gluon densities in the wavefunctions of the incoming hadrons, as
generated via QCD evolution (i.e.~parton branching) with increasing energy, and also --- 
in the case of a large nucleus with atomic number $A\gg 1$ --- via radiation from a large number 
of valence quarks. This paradigm has recently been challenged by new data for p+Pb
collisions at the LHC, which refer to long-range rapidity correlations (the `ridge effect')
and exhibit strong flow components  \cite{CMS:2012qk,Abelev:2012ola,Aad:2012gla,Aad:2013fja}, 
suggestive of `final-state interactions'
--- that is, interactions between the partons liberated by the collision and which might 
form a dense fireball at the intermediate stages of a collision. For nucleus-nucleus
collisions, the importance of final-state interactions is by now well established, via phenomena
like jet quenching or hydrodynamical flow, as observed at both RHIC and the LHC.
But even in that case, it is quite clear that the final multi-particle distribution as measured in the
detectors is the result of a complex interplay between initial-state and final-state interactions.
In order to distinguish between the various possible mechanisms and arrive at a global 
and unambiguous understanding of the data, it is essential and urgent to have reliable calculations of
multi-particle production, from first principles. 

The present study provides a further step in
that sense, by proposing a new and hopefully efficient method for computing multi-particle
production with rapidity correlations in a `dense-dilute' set-up, such as proton-nucleus collisions. 
Some physics problems that we have in mind are two-hadron production at central-forward 
rapidities\footnote{By `forward rapidities' we mean the fragmentation region of the dilute projectile.}
and the `ridge effect' alluded to above.
Our method includes the effects of  `initial-state interactions' in the framework of the 
Color Glass Condensate (CGC) effective theory 
\cite{Iancu:2003xm,Weigert:2005us,JalilianMarian:2005jf,Triantafyllopoulos:2005cn,Gelis:2010nm, Iancu:2012xa,Kovchegov:2012mbw}. 
In this framework, correlations in rapidity and transverse momentum (or azimuthal angle) 
are built in the nuclear wavefunction, via the JIMWLK evolution \cite{JalilianMarian:1997jx,JalilianMarian:1997gr,Kovner:2000pt,Weigert:2000gi,Iancu:2000hn,Iancu:2001ad,Ferreiro:2001qy} 
of the respective gluon distribution, and get transmitted to the produced partons via multiple scattering 
during the collision. Our approach generalizes previous CGC calculations of particle production 
which referred to simpler observables --- either single inclusive hadron production
\cite{Kovchegov:1998bi,Kovchegov:2001sc,Gelis:2002ki,Blaizot:2004wu,Blaizot:2004wv}, or a pair of
particles with similar rapidities (e.g.~di-hadron production at `forward rapidities', i.e.~in the 
fragmentation region of the projectile) 
\cite{Nikolaev:2003zf,JalilianMarian:2004da,Baier:2005dv,Marquet:2007vb,Albacete:2010pg,Dominguez:2011wm,Lappi:2012nh,Iancu:2013dta}.  
It builds up on previous extensions of the CGC formalism towards
multi-particle production \cite{JalilianMarian:2004da,Hentschinski:2005er,Kovner:2006ge,Kovner:2006wr}, 
which brought important conceptual clarifications but failed to provide a tractable
calculational scheme. In fact, the method that we shall develop here is the extension to particle
production of the only method that has proven so far to be useful for actually
solving the JIMWLK equation: the Langevin approach pioneered in Ref.~\cite{Blaizot:2002xy}
(see also Refs.~\cite{Rummukainen:2003ns,Lappi:2011ju,Dumitru:2011vk} for  
numerical implementations of this method).

The main difficulty with this kind of problems is the treatment of
the high-energy evolution at intermediate rapidities, between those of the 
produced particles. This cannot be factorized as `initial-state' (i.e.~prior to the collision) 
evolution of the (proton and nucleus) wavefunctions and greatly complicates
the calculation of the cross-sections.
To better explain this difficulty, we start with a problem for which the CGC factorization
is by now well established: the production of one or several partons with similar
rapidities in proton-nucleus \cite{Kovchegov:1998bi,Kovchegov:2001sc,Gelis:2002ki,Blaizot:2004wu,Blaizot:2004wv,Nikolaev:2003zf,JalilianMarian:2004da,Baier:2005dv,Marquet:2007vb,Albacete:2010pg,Dominguez:2011wm,Lappi:2012nh,Iancu:2013dta}, or nucleus-nucleus collisions \cite{Kovner:1995ts,Kovchegov:1997ke,Gelis:2008rw,Gelis:2008ad}.
The CGC factorization scheme, which holds within the leading logarithmic approximation 
in perturbative QCD at high energy, separates the high energy evolution of 
the incoming wavefunctions from the cross-section for the partonic subprocess, 
while taking into account the high-density effects (gluon saturation and multiple scattering) 
on the nuclear side.  
For definiteness, consider two gluon production in p+A collisions. So long as the
rapidity difference $\Delta\eta$ between the produced gluons is not too large, such that 
$\alpha_s\Delta\eta\ll 1$, one can ignore the intermediate evolution. 
It is then convenient to divide the total high-energy evolution between the BFKL evolution 
\cite{Lipatov:1976zz,Kuraev:1977fs,Balitsky:1978ic} of the dilute
projectile (the proton) and the JIMWLK evolution of the dense target (the nucleus), each
of them extending over the respective rapidity difference w.r.t. the produced gluons.
Moreover, when computing these evolutions, one finds that only the {\em initial-state} 
emissions contribute to the final result:  the effects of the {\em final-state} evolution
--- emissions of unresolved gluons which occur {\em after} the collision --- 
cancel between `real' and `virtual' emissions \cite{Chen:1995pa}. This is a consequence
of causality: within the approximations at hand, the evolution gluons are significantly
faster than the measured ones, so the respective formation times are also much
larger, by Lorentz time dilation. Hence, the final-state emissions associated with the
high-energy evolution occur long after the particle production has been completed
and cannot affect the cross-section for the latter.

We now move to the more interesting case where the rapidity separation between the
produced gluons is relatively large, $\alpha_s\Delta\eta\gtrsim 1$, and the intermediate evolution
cannot be neglected anymore. The (unresolved) gluons associated with this
evolution can in particular be emitted by the fastest among the two measured gluons, that is,
the one which is closer in rapidity to the projectile. Such
emissions can occur both before and after the scattering between the emitter and the target. 
Whenever such an emission occurs, it modifies the state (color and kinematics) of the fast measured 
gluon, hence it has an observable effect on the cross-section --- including for the final-state emissions.
Accordingly, the effects of the final-state evolution do not cancel anymore. Moreover these effects
depend upon the color field of the target: they describe the multiple scattering of the evolution
gluons off the strong target field. We see that the complications associated with the final-state
emissions are generally twofold: not only they cannot be simply factorized like the initial-state
evolution, but they can neither be treated as the standard BFKL evolution 
of the fast measured gluon from its own rapidity down to that of the softer measured
one\footnote{The second complication disappears in the limit where the target is
itself dilute and one can neglect multiple scattering. In that case, one recovers the
standard $k_T$ factorization of multi-particle production at high energy, based on
the BFKL evolution (see also the discussion in Appendix \ref{app-gfhier}).}.
Rather, one has to deal with BFKL evolution in the presence of a strong background field. 
A first method in that sense, which applies in the limit where
the number of colors $N_c$ is large and exploits the `color dipole' formulation  \cite{Mueller:1993rr}
of the BFKL evolution, has been proposed in \cite{JalilianMarian:2004da}. A more general method,
which works for any $N_c$ and involves a suitable generalization of the CGC formalism,
has been introduced in Ref.~\cite{Hentschinski:2005er}, in a somewhat different context
--- the study of diffraction. Subsequently, this method has been generalized to
particle production in \cite{Kovner:2006ge,Kovner:2006wr}. This second method will be here
given an equivalent Langevin formulation, which is better suited for numerical calculations.

The basic idea is that one needs to follow the background-field evolution of the wavefunction squared 
of the fastest measured gluon while keeping trace of the difference between the background field in the direct
amplitude (DA) and that in the complex conjugate amplitude (CCA). Indeed, at the intermediate stages 
of the calculations, this `background' field does not reduce itself to the classical field of the target
(which is of course the same in the DA and the CCA), but also includes the quantum fields
representing the softer gluons to be emitted --- the `evolution' gluons at intermediate rapidities
and the soft measured gluons. (This doubling of the quantum degrees of freedom is reminiscent
of the Keldysh-Schwinger formulation of quantum field theory in real time, which involves
a similar doubling of the time axis, with the upper side corresponding to the DA,
and the lower side to the CCA.) Hence, the wavefunction squared plays the role of a {\em generating
functional} for both evolution and particle production. Its evolution can be viewed from either the
projectile side, or from the target side, and it is the last perspective which will be more
useful for us here: as in the standard CGC formalism, it is the target perspective which allows 
for a Langevin reformulation of the JIMWLK evolution \cite{Blaizot:2002xy}.

To describe our method, let us first remind that, in a Lorentz frame where the target carries most
of the high-energy evolution --- a `target infinite momentum frame' ---, the partons from the projectile couple
to the (strong) color field of the target via {\em Wilson lines}, i.e.~time-ordered exponentials of the
gauge potential, where the `time' is the {\em light-cone} time of the {\em projectile} (i.e.~$x^+$ if the
projectile is a right-mover). The time ordering of the fields in the Wilson lines reflects their
ordering in rapidity: the inner core near the light-cone ($x^+=0$) corresponds to the target 
valence degrees of freedom and is surrounded by layers of fields at larger and larger values
of $|x^+|$, associated with quantum gluons which are softer and softer relative to the target
(i.e.~closer and closer to the rapidity of the projectile). This `multi-layer' structure is manifest in the
Langevin formulation of the JIMWLK evolution, where the Wilson lines are built by successively adding 
new layers at larger values of $|x^+|$ with increasing the rapidity difference
w.r.t. to the (valence d.o.f. of the) target. 
In the original formulation of the CGC effective theory, which is oriented towards
the calculation of the elastic $S$-matrix for dilute-dense scattering, all the modes of the target field
are treated as {\em classical} fields, irrespective of their rapidity. This formulation can also be used
for computing multi-particle production, but only so long as the produced partons have similar
rapidities. In that case, one can associate by hand two Wilson lines --- one in the DA, the other one
in the CCA --- to each of the produced partons. These two Wilson lines live at different transverse coordinates,
but they are both built with the same gauge field --- the classical field of the target. 
However this construction becomes insufficient for the general problem of multi-particle production, 
where the measured partons have arbitrary rapidities. 
In that case, the measured gluons have to be put on-shell, hence they must be treated
as {\em quantum} fields, which are different in the DA and respectively the CCA. In the presence
of non-linear phenomena like gluon saturation and multiple scattering, this distinction must be
carried over to all the `evolution' (unresolved) gluons at intermediate rapidities, in between the
produced ones. This in turn requires a generalized CGC formalism, in which Wilson lines in the 
DA and the CCA are built independently from each other.

The simultaneous evolution of the target fields in the DA and the CCA is governed by a generalization
of the JIMWLK Hamiltonian \cite{Hentschinski:2005er} --- essentially, its extension to the
Keldysh-Schwinger closed time contour --- that we shall refer to
as the `evolution Hamiltonian' $H_{\rm evol}$. In Refs.~\cite{Kovner:2006ge,Kovner:2006wr},
this has been mostly employed to study the evolution of the (dilute) projectile at large $N_c$, and thus deduce 
equations which extend the Balitsky-Kovchegov equation \cite{Balitsky:1995ub,Kovchegov:1999yj}
to the cross-section for two gluon production\footnote{These equations are
equivalent to those constructed in \cite{JalilianMarian:2004da} from the dipole picture.}. 
This strategy will be illustrated 
in Appendix \ref{app-gfhier}, where we shall use $H_{\rm evol}$ to construct coupled evolution 
equations (valid for any $N_c$), which generalize the Balitsky hierarchy \cite{Balitsky:1995ub}
to multi-particle production in dense-dilute scattering. The ensuing equations are however extremely 
complicated, both because of the non-linear effects associated with saturation in the target, and 
because of the transverse non-locality inherent in the calculation of particle production.
Some simplifications occur at large $N_c$, where the hierarchy closes at the level of the
`dipole' and the `quadrupole' --- color traces of two and, respectively, four Wilson lines
in the fundamental representation, which in the present context represent generating 
functionals for projectiles made with a single quark and, respectively, a quark-antiquark dipole.
But even those equations appear to be too complicated to be useful in practice. (The simplest
closed equation, which describes gluon radiation by a color dipole,
has the same degree of complexity as the Balitsky-JIMWLK equation for the evolution of
the quadrupole $S$-matrix
\cite{JalilianMarian:2004da,Kovner:2006ge}; see \eqn{nk2evol} in Appendix \ref{app-gfhier}.)

This complexity appears to be inextricable and motivates our search for an alternative strategy.
Our proposal is to use  $H_{\rm evol}$ in the same way as the original JIMWLK Hamiltonian is
generally used in the context of the CGC: to describe the evolution of the {\em target}. This 
allows us to set-up a Langevin formalism in which the Wilson lines in the DA and respectively
the CCA are built independently from each other, over the whole rapidity interval separating
the produced particles. One additional difficulty as compared to the standard JIMWLK evolution
is the fact that, as alluded to above, one needs to construct a {\em generating functional} ---
a functional of the Wilson lines at lower values of the rapidity, which can be used
to generate an arbitrary number of gluons via functional differentiations. In terms
of the Langevin process, this implies that one has to solve a stochastic equation with {\em
functional} initial conditions (generic Wilson lines, rather than numerically-valued color matrices).
This is conceptually well-defined, but not well suited for numerics. The solution that we
propose to circumvent this difficulty, is to adapt the Langevin process to the specific cross-section of
interest: for a given final state, one can build {\em ordinary} (i.e.~non-functional) Langevin equations
for both the Wilson lines and their functional derivatives which enter the calculation of the cross-section.
For instance, in order to compute two gluon production with generic rapidity separation, one needs
additional Langevin equations for the first-order functional derivatives of the Wilson lines, 
for three gluon production, one also needs the equations obeyed by the respective 
second-order functional derivatives, etc. All such equations are straightforward to write down and 
{\em a priori} accessible to numerical simulations. But the complexity of the numerical implementation
should rapidly increase with the number of produced particles: indeed, each additional functional
derivative increases the non-locality of the Langevin process in the transverse space
(see Sect.~\ref{sect-langevin} for details). Hopefully, this whole scheme will turn out to 
be tractable, at least, for the calculation of two-gluon production, with
consequences for the phenomenology of di-hadron correlations in proton-nucleus collisions 
at RHIC and the LHC.

As mentioned earlier, the physical original of the large ridge effect seen in p+Pb collisions at the LHC 
is still unclear. Previous CGC-inspired calculations appear to describe these data quite well 
\cite{Dusling:2012cg,Dusling:2012wy,Dusling:2013oia} (see also 
Refs.~\cite{Dumitru:2008wn,Gelis:2008sz,Dumitru:2010iy} for earlier related work), 
but this agreement cannot be 
viewed as fully conclusive in view of the many underlying approximations, whose effect is difficult to
control. We hope that the method proposed in this paper will open the way towards controlled 
calculations from first principles, within the accuracy limits of the present formalism.

The plan of the paper is as follows. In Sect.~\ref{sect-CGC} we give a brief review of the
CGC formalism and the JIMWLK equation, with emphasis on the Langevin formulation
of the latter.  In Sect.~\ref{sect-same} we consider multi-particle production at similar
rapidities, that is, in the absence of any high energy evolution between the parent
partons and the produced ones. In this context, we introduce the 
notion of generating functional for particle production (the `wavefunction squared of the
projectile'), together with a special operator --- the `production Hamiltonian' $H_{\rm prod}$ --- 
which generates on-shell gluon emissions when acting on the generating 
functional \cite{Kovner:2006ge,Kovner:2006wr}. 
This Hamiltonian looks very similar to the JIMWLK Hamiltonian, for reasons which should become 
clear in the next section. Specifically, in Sect.~\ref{sect-different} we study the high energy evolution 
with increasing rapidity difference between the parent partons and the produced gluons, 
or between the produced gluons themselves. As already mentioned, this evolution
is governed by $H_{\rm evol}$ 
--- a generalization of the JIMWLK Hamiltonian which generates gluon emissions in both the DA and the CCA \cite{Hentschinski:2005er}.
(The `production Hamiltonian' $H_{\rm prod}$ is the `cut' version of $H_{\rm evol}$, which
produces on-shell gluons alone.) The emphasis in this section will be on the evolution
of the target (the dense nucleus), up to the rapidity of the `most forward' produced parton. 
(The complimentary viewpoint of the projectile evolution will be developed in Appendix \ref{app-gfhier}.)
In this context, we shall describe the generalization of the CGC formalism to multi-particle production. 
This involves an `off-diagonal' version of the CGC weight function, which allows for different 
field configurations in the DA and, respectively, the CCA, and encodes the high-energy 
evolution of the generating functional. Sect.~\ref{sect-langevin} presents our main new
results: a Langevin formulation for the JIMWLK evolution of the generating functional.
In particular, we shall present a version of the formalism which is free of functional
aspects and thus suitable for numerical implementations. In Sect.~\ref{sect-langevin}, we show 
the stochastic equations appropriate for two gluon production, while in Appendix \ref{app-three} 
we describe the additional equations which appear when computing three gluon production 
with generic rapidity differences.

\section{The Color Glass Condensate and the JIMWLK evolution}
\label{sect-CGC}

In the effective theory of the Color Glass Condensate (CGC) valid in the leading logarithmic
approximation at high energy, the expectation value of an observable $\hat{\mcal{O}}$ which 
is local in rapidity and which is associated with the scattering between a dilute projectile (`proton') and 
a dense target (`nucleus'), is schematically given by the following functional integral
 \begin{align}
 \label{oave}
 \langle \hat{\mcal{O}} \rangle_Y = \int [DU]\, W_Y[U]\, \hat{\mcal{O}}.
 \end{align}
Here, $U\equiv U({\bx})$ is the Wilson line 
describing the scattering between a parton with transverse coordinate $\bx$ from the projectile and
the strong color field of the target, in the eikonal approximation (see  \eqn{Udef} below
for an explicit formula). $[DU]$ is the functional group-invariant (or de Haar) measure on SU$(N_c)$.
$W_Y[U]$ is a functional probability density describing the distribution of the Wilson lines in the
target, known as the  `CGC weight-function'.
$Y$ is the rapidity difference between the projectile and the target,
 and it is assumed to be large, that is, $\alpha_s Y > 1$ with $\alpha_s$ the QCD coupling. 
 \eqn{oave} is written in a frame in which most of this rapidity is carried by the target, 
so one can ignore the high-energy evolution of the projectile.
For instance\footnote{From now on, we shall denote the dependence of the Wilson lines upon the
transverse coordinates with an index instead of an argument.},
 \begin{align}
 \label{sgdip}
 \hat{S}(\bx\by) = \frac{1}{N_g}\,
 \rmTr[U_{\by}^\pd U^{\dagger}_{\bx}], 
 \end{align}
represents the $S$--matrix for the scattering of a right-moving gluonic dipole 
off the strong color field of the left-moving nucleus. In the above $N_g = N_c^2 -1$ 
with $N_c$ the number of colors, and
 \begin{align}\label{Udef}
 U^{\dagger}_{\bx} = \rmP \exp\left[\rmi g \int \dif x^+ \alpha^a_{\bx}(x^+) T^a\right],
 \end{align}
where the $T^a$'s are the color group generators in the adjoint representation, 
$A_a^\mu=\delta^{\mu -}\alpha_a$ is the color field generated by the nucleus 
and P stands for path ordering: with increasing $x^+$, 
matrices are ordered from right to left. The integral over $x^+$ formally extends along the real axis,
but in practice it is limited to the support of the target field, which is localized near $x^+=0$ by Lorentz
contraction (a `shockwave'). This support depends upon the rapidity difference $Y$ : with increasing $Y$, 
the scattering probes gluon modes in the nuclear wavefunction which carry smaller and smaller fractions
$x=\rme^{-Y}$ of the target longitudinal momentum ($k^-$),
and hence are more and more delocalized in $x^+$. 
The information about the support of the target field and about its correlations
is encoded in the CGC weight-function $W_Y[U]$. When increasing $Y$, this
evolves according to a functional renormalization group equation known as the JIMWLK equation\footnote{The acronym stands for Jalilian-Marian, Iancu, McLerran, Weigert, Leonidov and Kovner.} \cite{JalilianMarian:1997jx,JalilianMarian:1997gr,Kovner:2000pt,Weigert:2000gi,Iancu:2000hn,Iancu:2001ad,Ferreiro:2001qy}  :
 \begin{align}
 \label{rg}
 \frac{\del W_Y[U]}{\del Y} = H W_Y[U].
 \end{align}
$H$ is the JIMWLK Hamiltonian, which for our purposes is most conveniently written as\footnote{We shall use in general the shorthand notation
$\int_{\bu\bv \dots}$ instead of $\int \dif^2 \bu\, \dif^2 \bv \dots$.}
 \begin{align}
 \label{jimwlk}
 H = \frac{1}{8\pi^3}
 \int_{\bu\bv\bz}
 \mcal{K}_{\bu\bv\bz}
 \big[L^a_{\bu} - U^{\dagger ab}_{\bz} R^b_{\bu}\big]
 \big[L^a_{\bv} - U^{\dagger ac}_{\bz} R^c_{\bv}\big],
 \end{align}
 where $\mcal{K}_{\bu\bv\bz}
 \equiv \mcal{K}^i_{\bu\bz} \mcal{K}^i_{\bv\bz}$ with $\mcal{K}^i_{\bu\bz}$ 
 the Weizs\"{a}cker-Williams emission kernel:
   \begin{align}
 \mathcal{K}^i_{\bu\bz} = \frac{(\bu - \bz)^i}{(\bu - \bz)^2}.
 \end{align} 
In \eqn{jimwlk},
$L$ and $R$ are `left' and `right' Lie derivatives, i.e.~the generators of local color rotations 
of the Wilson lines on the left and, respectively, on the right. They can be formally, but unambiguously, defined via their action on the Wilson lines, which reads
 \begin{align}\label{Lie}
 L^a_{\bu} U^{\dagger}_{\bx}= 
 \rmi g \delta_{\bu\bx} T^a U^{\dagger}_{\bx}, \qquad
 R^a_{\bu} U^{\dagger}_{\bx}= 
 \rmi g \delta_{\bu\bx} U^{\dagger}_{\bx} T^a.
 \end{align}
For the purpose of the physical interpretation, it is however useful to notice that the Lie
derivatives can be explicitly realized as functional derivatives with respect to the field
$\alpha^a_{\bx}(x^+)$ at the end-points of the Wilson lines: the $L$ derivative acts at the 
largest value of $x^+$ (which is positive), meaning {\em after} the scattering with the shockwave, 
whereas the $R$ derivative acts at the smallest value of $x^+$ (which is negative), 
meaning {\em before} the scattering. Hence, each step in the evolution described by
Eqs.~\eqref{rg} and \eqref{jimwlk} adds two infinitesimal layers, one `to the left'  and  one `to the right', 
to the support of the target field: with increasing $Y$ the shockwave expands
in $x^+$, as anticipated. This expansion is however not symmetric under reflexion in $x^+$,
as shown by the structure of the JIMWLK Hamiltonian, where the Wilson lines
multiply the {\em right} derivatives alone.
The physical meaning of this dissymmetry will be explained later.

The Lie derivatives are non-commuting objects which obey the color group algebra (with $f^{abc}$ the
structure constants for SU$(N_c)$)
\begin{align}\label{Liecom}
[ L^a_{\bu},\, L^b_{\bv}]= g \delta_{\bu\bv} f^{abc}L^c_{\bu}\,,\qquad
[ R^a_{\bu},\, R^b_{\bv}]= -g \delta_{\bu\bv} f^{abc}R^c_{\bu}\,,
\end{align}
yet one does not need to worry about the particular ordering of the derivatives in \eqn{jimwlk} since
 \begin{align}\label{comm}
 [L^a_{\bu},R^b_{\bv}] = 
 [L^a_{\bu},U^{\dagger ab}_{\bz}]=
 [R^a_{\bu},U^{\dagger ab}_{\bz}]=0.
 \end{align}
Note also that the `left' and `right' Lie derivatives are not independent operators, rather they
are related by a color rotation with the Wilson line: $L^a_{\bu}= U^{\dagger ab}_{\bu}R^b_{\bu}$.
This relation is a consequence of the unitarity of the Wilson lines and can be checked by using
\eqn{Lie} together with the color group identity $U^{\dagger ab} T^b = U T^a U^{\dagger}$.
 
We are interested in the evolution of the observables with increasing $Y$. After taking
a derivative with respect to $Y$  in \eqn{oave} and using \eqn{jimwlk}, there are two ways to proceed. The first one
is to integrate by parts the Lie derivatives and make them act on the operator $\hat{\mcal{O}}$  which
defines the observable. This amounts to transferring the evolution step from the wavefunction of the
target to that of the projectile. From this perspective, the role of the Lie derivatives is to generate 
soft gluon emissions from the color sources represented by the Wilson lines.
This procedure yields an evolution equation for $\langle \hat{\mcal{O}} \rangle_Y$, 
which however is not a closed equation, but a member in an 
infinite hierarchy of coupled equations for the correlations of the Wilson lines --- 
the Balitsky hierarchy \cite{Balitsky:1995ub}. This hierarchy becomes tractable in the limit of large $N_c$, where 
the first respective equation reduces to a closed, non-linear, equation --- the Balitsky-Kovchegov equation
\cite{Balitsky:1995ub,Kovchegov:1999yj}. The solution to this equation can be combined with
appropriate truncation schemes, like a Gaussian approximation
\cite{Iancu:2002aq,Blaizot:2004wv,Kovchegov:2008mk,Marquet:2010cf,Dominguez:2011wm,Iancu:2011ns,Iancu:2011nj,Alvioli:2012ba}, and used to construct solutions for the higher equations in the hierarchy. 
The second strategy (the only one to be useful at finite $N_c$ and whenever the Gaussian approximation 
fails to apply),  is based on the fact that Eqs.~\eqref{rg} and \eqref{jimwlk} correspond to a 
functional Fokker-Planck equation, that can be given an equivalent Langevin formulation \cite{Blaizot:2002xy}, 
which is better suited for numerical studies \cite{Rummukainen:2003ns,Lappi:2011ju,Dumitru:2011vk}.

\begin{figure}
\begin{center}
\includegraphics[width=0.32\textwidth,angle=0]{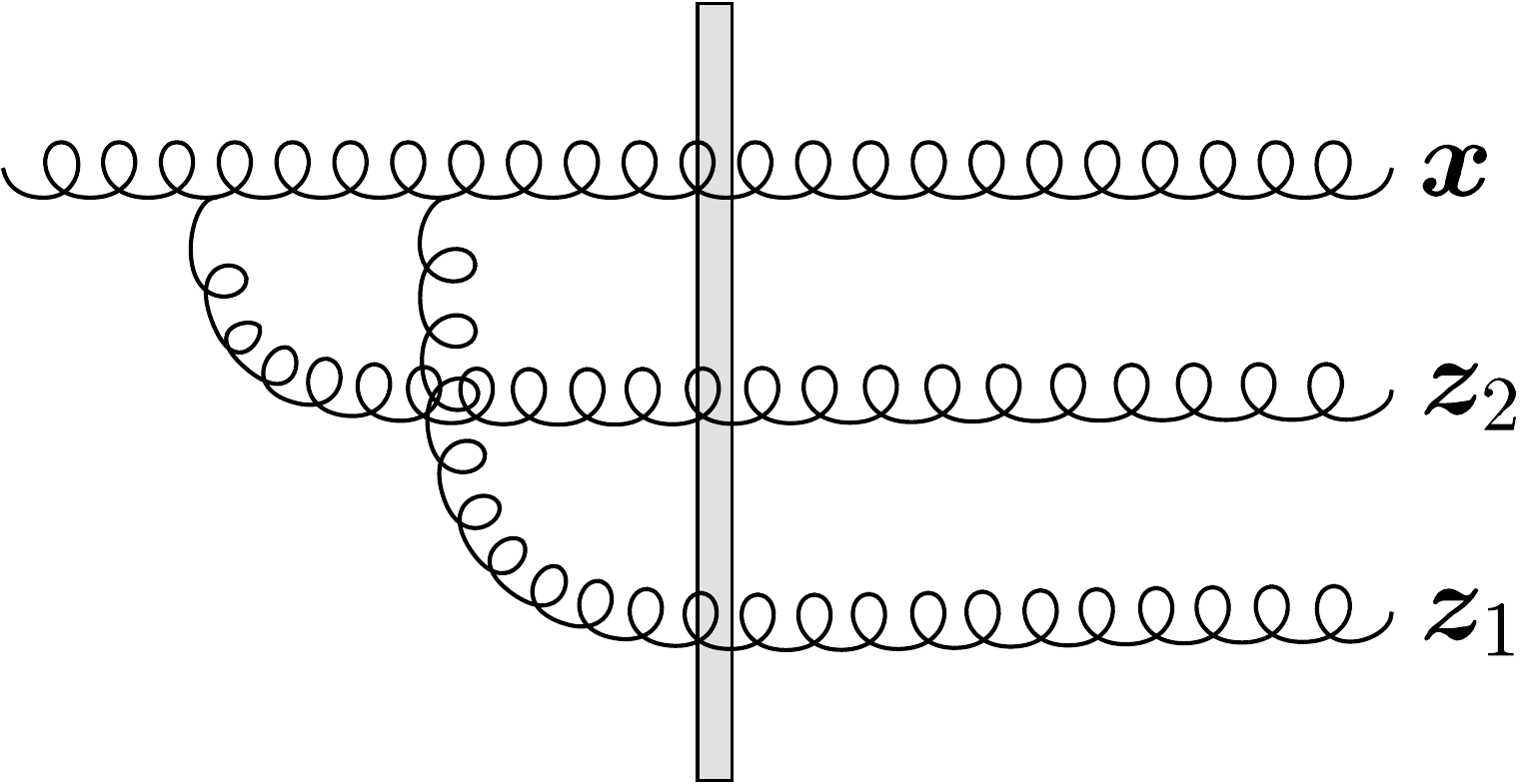}\,
\includegraphics[width=0.32\textwidth,angle=0]{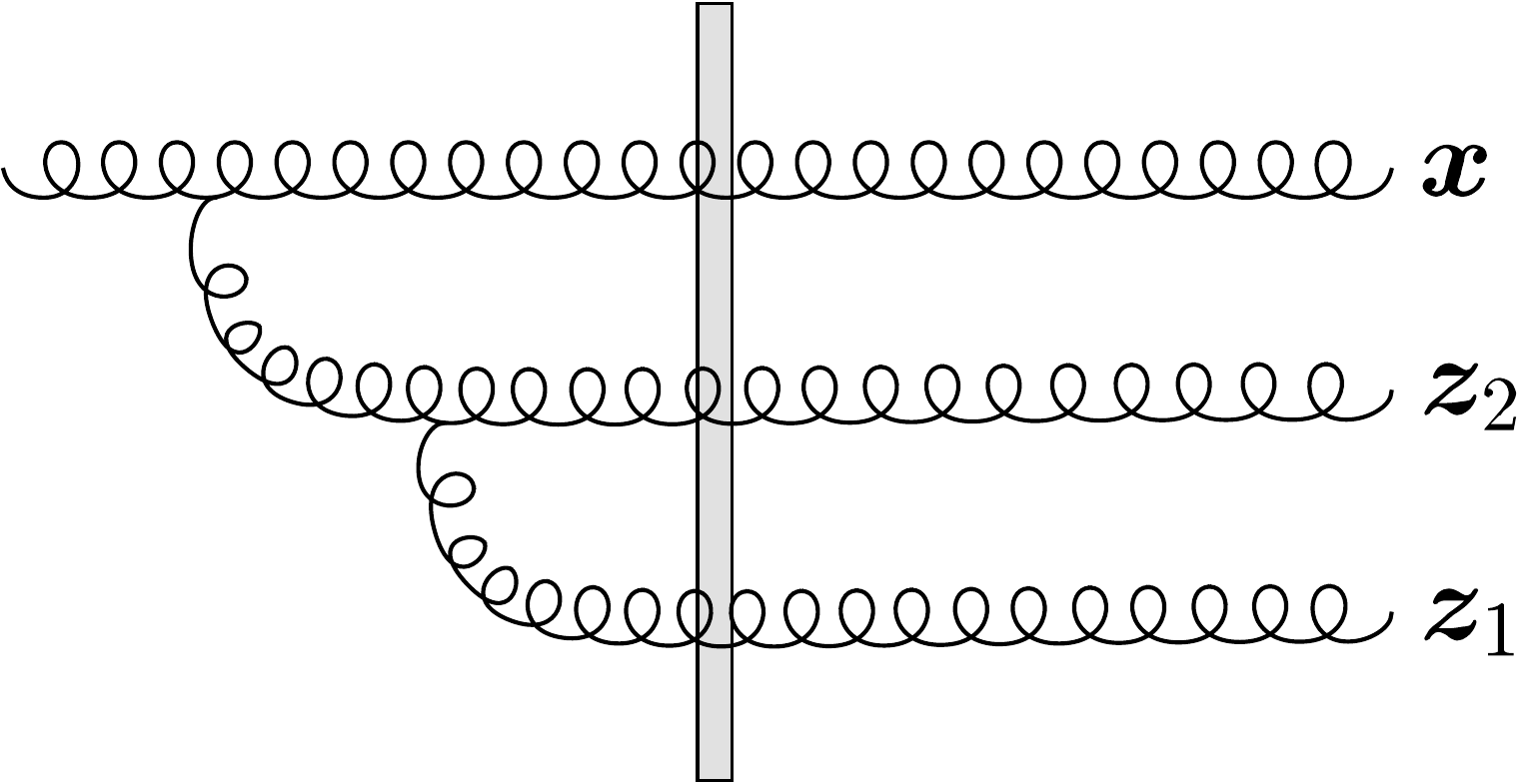}\,
\includegraphics[width=0.33\textwidth,angle=0]{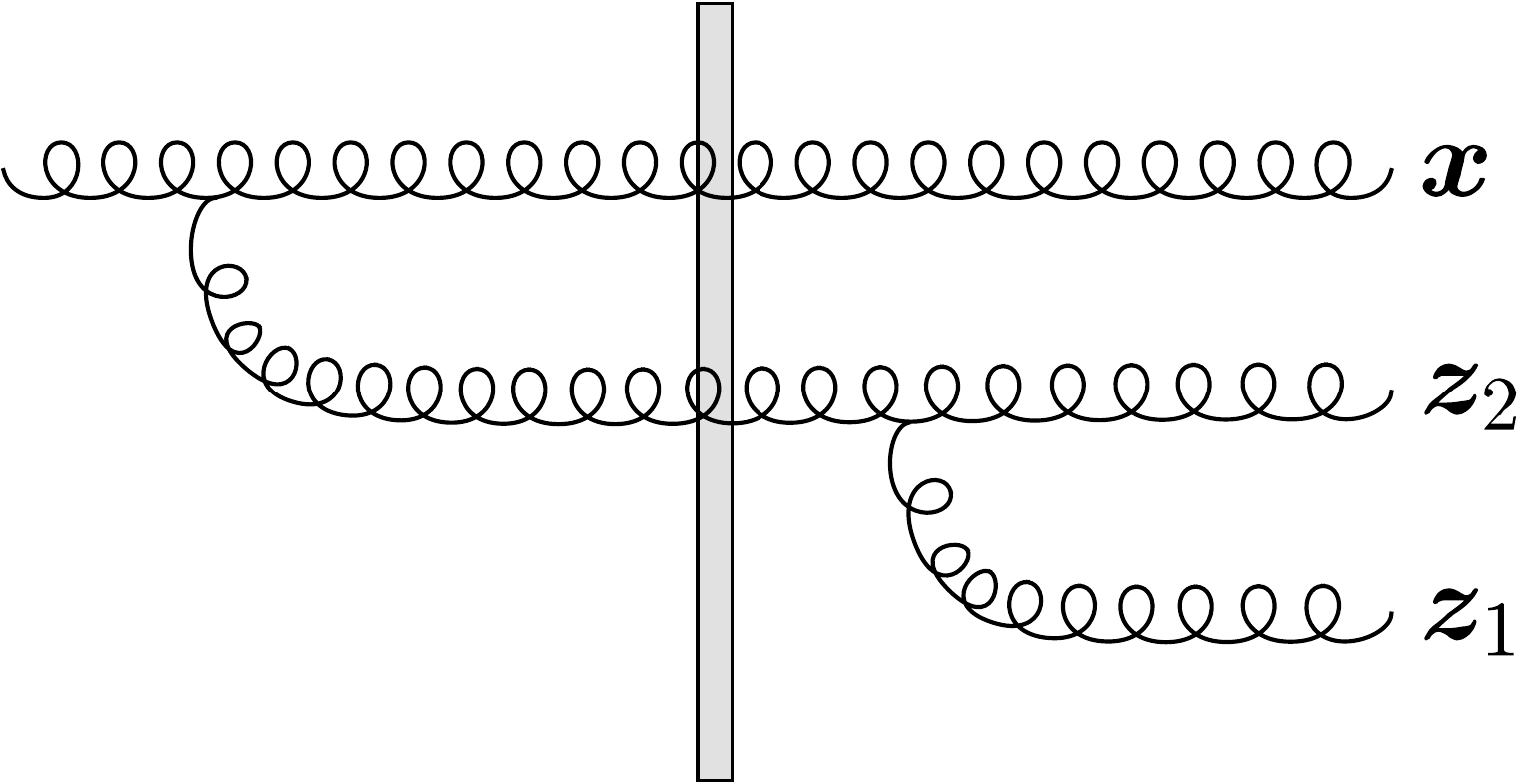}\,
\end{center}
\caption{Two successive steps in the evolution described by \eqn{ulang}. Gluon `2' has a larger $k^+$ than gluon `1' ($k_2^+ > k_1^+$), but a smaller $k^-$ ($k_2^- < k_1^-$). From the viewpoint of target evolution, cf.~\eqn{ulang}, gluon `2' is emitted after gluon `1', so in particular it can scatter off the color field created by the latter. From the dual viewpoint of projectile evolution, gluon `2' is emitted first and can act as a source for gluon `1'. The vertical line represents the shockwave of the nucleus, located near $x^+=0$.}
\label{fig:twosteps}
\end{figure}

In the Langevin formulation, the JIMWLK evolution of the color field in the target is depicted as a random walk in the 
functional space of the Wilson lines, with $Y$ playing the role of `time'.  The CGC average of any observable,
in the sense of \eqn{oave}, can be then computed as an average over the noise term in the 
Langevin equation which governs this random walk. For instance, for the gluonic dipole in \eqn{sgdip}, one writes
 \begin{align}
 \label{sgdipave}
 \big\langle \hat{S}_{\bx\by} \big\rangle_Y = 
 \frac{1}{N_g}
 \big\langle \rmTr[U_{N,\by}^\pd U_{N,\bx}^{\dagger}] \big\rangle_{\nu}\,,
 \end{align}
where we have discretized the rapidity interval according to $Y-Y_{\rm in}=\epsilon N$, with $N \to \infty$ and $\epsilon \to 0$. (The `initial' rapidity $Y_{\rm in}$ is where we introduce the initial conditions for the evolution;
see below.) As already mentioned, each evolution step adds two new layers in the support
of the Wilson lines in $x^+$, leading to infinitesimal,
left and right, color rotations. These rotations are random, reflecting the quantum nature of the
fluctuations that have been `integrated' out. This leads to the following Langevin equation
 \cite{Iancu:2011nj,Lappi:2012vw}
 \begin{align}
 \label{ulang}
 U^{\dagger}_{n,\bx} = \exp[\rmi \epsilon g \alpha^{L}_{n,\bx}]\,
 U^{\dagger}_{n-1,\bx} \exp[-\rmi \epsilon g \alpha^{R}_{n,\bx}],
 \end{align}
where the left and right color matrix fields read
 \begin{align}
 \label{alphal}
 &\alpha^{L}_{n,\bx} = 
 \frac{1}{\sqrt{4\pi^3}}\int_{\bz} \mcal{K}^i_{\bx\bz} \,\nu^{ia}_{n,\bz}\, T^a,
 \\
 \label{alphar}
 &\alpha^{R}_{n,\bx} = 
 \frac{1}{\sqrt{4\pi^3}} \int_{\bz} \mcal{K}^i_{\bx\bz} \,\nu^{ia}_{n,\bz} U^{\dagger ab}_{n-1,\bz}\, T^b.
 \end{align}
In Eqs.~\eqref{alphal} and \eqref{alphar} $\nu_n$ is a Gaussian white noise local in rapidity, that is
 \begin{align}
 \label{noise}
 \big\langle \nu^{ia}_{m,\bx}\, \nu^{jb}_{n,\by} \big\rangle
 = \frac{1}{\epsilon}\,\delta^{ij} \delta^{ab} \delta_{mn} \delta_{\bx\by},
 \end{align} 
and this is the meaning of the average over $\nu$ in \eqn{sgdipave}. Physically, the noise term accounts for
the color charge density and the polarization of the target gluons radiated in this particular 
step of the evolution,  which act as sources for the new layers $\alpha^{R}_{n}$ and $\alpha^{L}_{n}$
in the target field.
These sources too are left movers, but they are not as fast as the sources produced in the
previous steps. Accordingly, the gluons radiated at negative $x^+$, meaning
{\em ahead} of the shockwave, can be caught by the latter, and then they are color--rotated. This is the origin 
of the Wilson line visible in the r.h.s.~of \eqn{alphar}, which in turn is responsible for generating the BFKL
cascade via iterations. Alternatively, and perhaps simpler, one can interpret \eqn{ulang} as
one step in the {\em projectile}
evolution; then, $\nu$ represents an on-shell gluon radiated by the projectile, 
which is a right mover and hence it can scatter off the target field, provided it was emitted at negative $x^+$
(i.e.~prior to the scattering).
Fig.~\ref{fig:twosteps} illustrates two steps in the evolution generated by \eqn{ulang}.

For the purpose of computing correlation functions, or deriving the associated evolution equations, one needs
to keep terms up to order $\epsilon$ in the r.h.s.~of the Langevin equation  \eqref{ulang}. Note however
that the noise term scales like $1/\sqrt{\epsilon}$, as manifest on \eqn{noise}, so in order to achieve the desired 
accuracy, one has to expand the left and right rotations in \eqn{ulang} up to quadratic terms. Moreover, the quadratic terms can already be replaced by their average, since they cannot be multiplied by  noise  terms
coming from other sources to the order of interest. When doing that, the
quadratic terms coming from the right rotation become independent of the Wilson line in the previous step (this is trivially true for left rotations) and correspond to virtual terms. Indeed one finds
 \begin{align}
 \label{rightquad}
 -\frac{\epsilon^2 g^2}{2}\, \big\langle (\alpha^{R}_{n,\bx})^2 \big \rangle &= 
 -\frac{\epsilon^2 g^2}{8\pi^3}\int_{\bz\bw}
 \mcal{K}^i_{\bx\bz} \mcal{K}^j_{\bx\bw}
 U^{\dagger ab}_{n-1,\bz}  U^{\dagger cd}_{n-1,\bw} T^b T^d
 \big\langle \nu^{ia}_{n,\bz}\, \nu^{jc}_{n,\bw} \big\rangle  \nn & =
 -\frac{\epsilon\abar }{2\pi} \int_{\bz} \mcal{K}_{\bx\bx\bz}\,=\, 
 -\frac{\epsilon^2 g^2}{2}\, \big\langle (\alpha^{L}_{n,\bx})^2 \big \rangle
 \end{align}

In both approaches to the JIMWLK evolution (the Balitsky hierarchy and the Langevin formulation),
one needs an initial condition, that is, the CGC weight-function at a given rapidity $Y_{\rm in}$. 
The initial condition is typically given by the MV (McLerran-Venugopalan) model \cite{McLerran:1993ni,McLerran:1993ka} in which the color charge $\rho$ of the nucleus is distributed according to a Gaussian distribution. Then, in principle, one can construct the initial values for all the correlators that appear in the Balitsky hierarchy.
In the Langevin approach, a Gaussian initial condition means that the color charges are randomly distributed on the two-dimensional transverse plane and, in turn, this determines a probability distribution for the initial values of $U_{\rm in}$ and $U_{\rm in}^{\dagger}$ needed to start developing the Wilson lines according to \eqn{ulang}.

\section{Generating functional and same rapidity gluon production}
\label{sect-same}

So far, the gluonic dipole introduced in \eqn{sgdip} has been interpreted as a {\em scattering amplitude},
namely the amplitude for the elastic scattering between a pair of gluons in a color singlet state
and the strong color field of the target. But the same quantity also enters the {\em cross-section} for 
the inclusive production of a gluon with transverse momentum $\bp$ and pseudo-rapidity $\eta_p$
in proton-nucleus collisions:
 \begin{align}
 \label{ptbroad}
 \frac{\dif \sigma_{1g}}{\dif \eta_p\dif^2 \bp} =\,xg(x)\,
 \frac{1}{(2\pi)^2} \int_{\bx\bbx} \rme^{-\rmi \bp \cdot (\bx - \bbx)} 
\langle \hat{S}_{\bx\bbx}  \rangle_Y,
\end{align}
where $xg(x)$ is the usual (`integrated') gluon distribution in the proton and $x$ is the longitudinal
momentum fraction of the projectile gluon participating in the scattering. (Note that we assume
collinear factorization at the level of the proton.) 
In this context, the Fourier transform in \eqn{ptbroad} describes the transverse momentum 
broadening of a gluon preexisting in the projectile. The two Wilson lines within $\hat{S}_{\bx\bbx}$ now describe
the same gluon, which has transverse coordinate $\bx$ in the direct amplitude (DA) and respectively $\bbx$ 
in the complex conjugate amplitude (CCA). The color trace and the normalization factor
$1/N_g$ have been generated by the sum (average)
over the gluon color indices in the final (initial) state.

In what follows, we will be mostly interested in
the radiation of new gluons triggered by the interactions between the projectile 
--- that will be taken to be simply a gluon --- and the target.
To compute the corresponding cross-section, it is convenient to introduce a generalization 
of the `gluonic dipole' in \eqn{sgdip}, which distinguishes between the Wilson lines in the DA
(for which we shall use the same notations as before, i.e.~$U$ and $U^{\dagger}$) and those
in the CCA (to be denoted with a bar: $\bar{U}$ and $\bar{U}^\dagger$). Specifically,
we introduce the following functional of $U$ and $\bar{U}$  
  \begin{align}
 \label{ggf}
 \hat{S}_{12}(\bx\bbx) = \frac{1}{N_c}\, \rmTr[\bar{U}_{\bbx}^\pd U^{\dagger}_{\bx}],
 \end{align}
which although a mathematical dipole, it physically describes a single gluon. 
\eqn{ggf} should be thought of as the `wavefunction squared' (WFS) of the gluonic projectile: 
it acts as a generating functional for radiation from this gluon in the presence of the nuclear shockwave.

\begin{figure}
\begin{center}
\includegraphics[width=0.45\textwidth,angle=0]{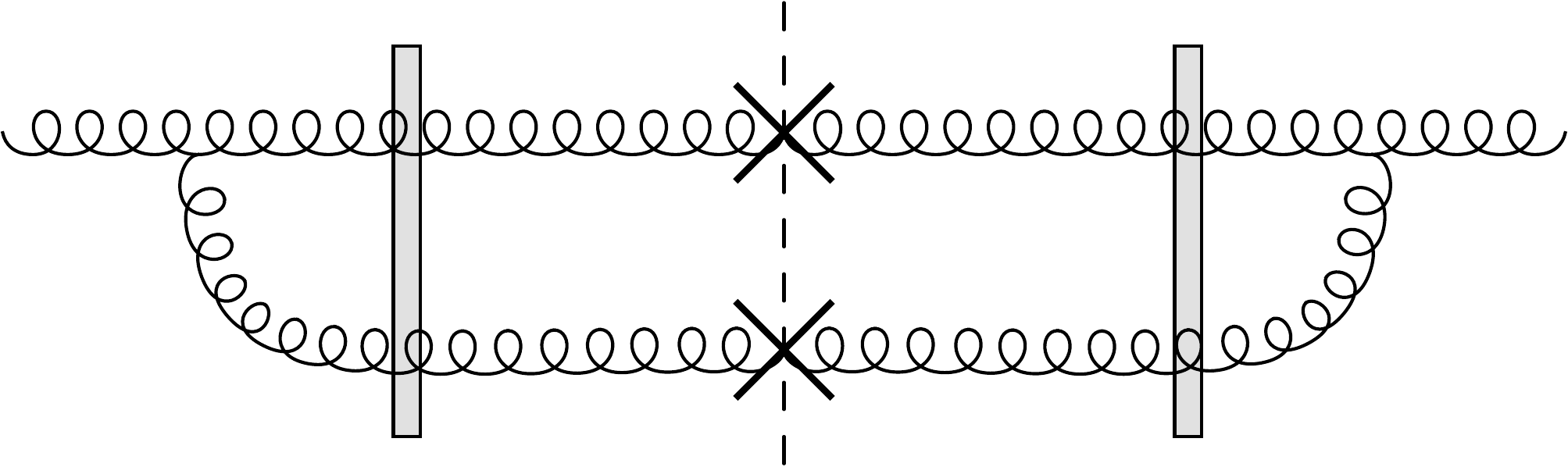}\qquad
\includegraphics[width=0.45\textwidth,angle=0]{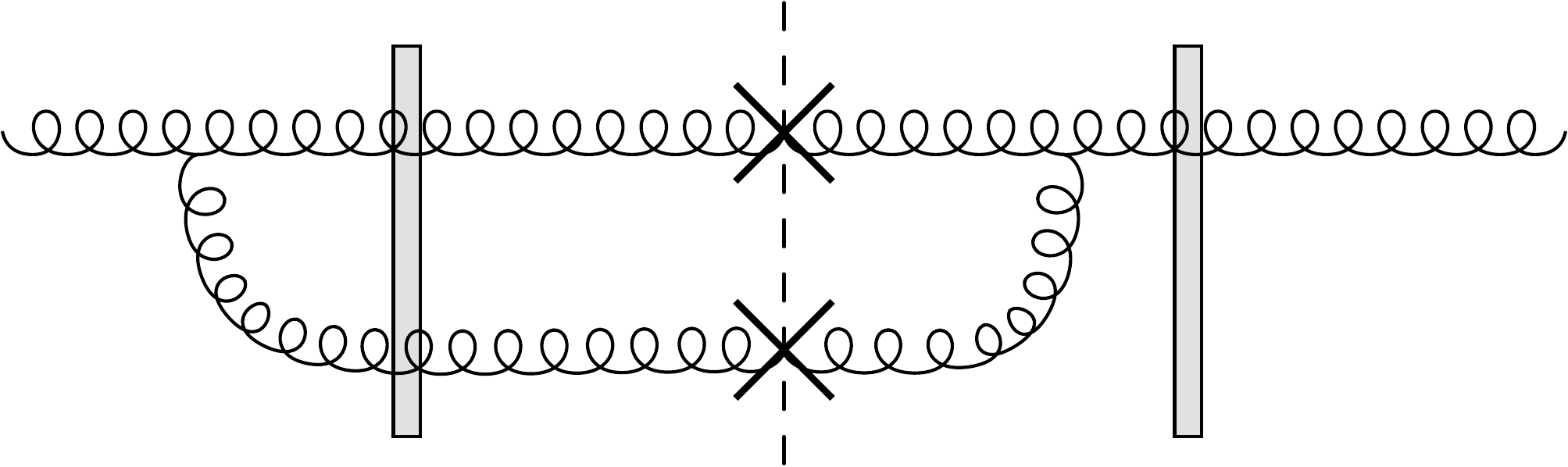}
\end{center}
\begin{center}
\includegraphics[width=0.45\textwidth,angle=0]{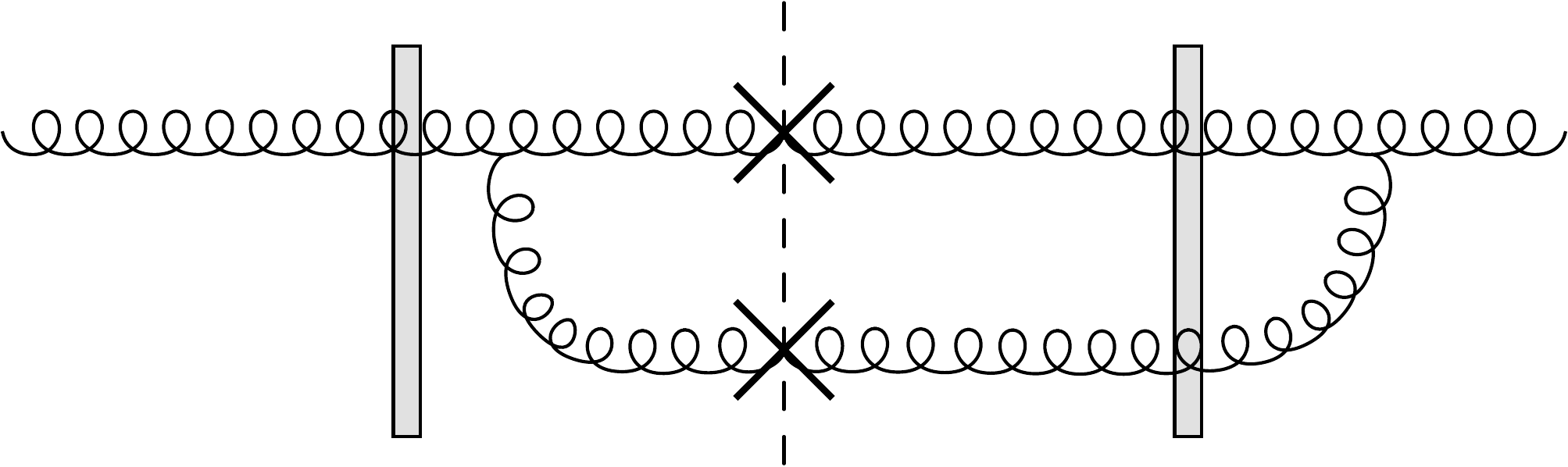}\qquad
\includegraphics[width=0.45\textwidth,angle=0]{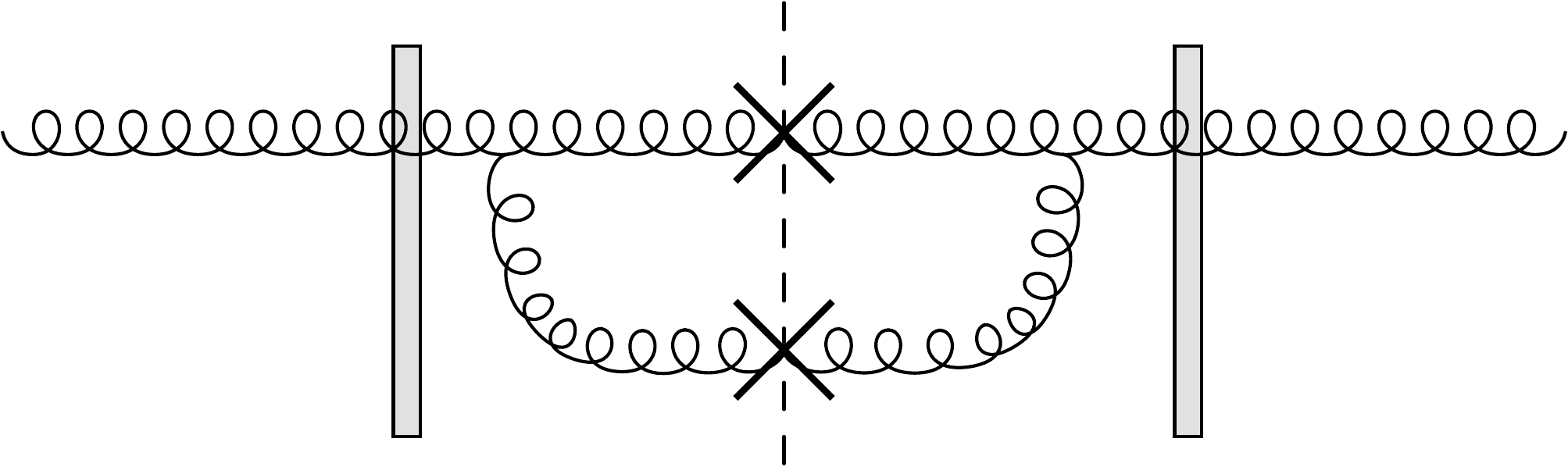}
\end{center}
\caption{Diagrams for the production of two gluons at the same rapidity. A cross stands for each gluon produced.}
\label{fig:twogluons}
\end{figure}

The above should become clear via examples.
To start with, consider the emission of a second gluon which is softer than the first one, but not {\em much}
softer: the emission vertex can be described in the eikonal approximation, but the
rapidity separation between the two gluons remain small enough to allow one to neglect
the high-energy evolution in between. In that sense both gluons have the same rapidity 
difference $Y$ w.r.t.~the target. Let ($\eta_p,\bp$) and ($\eta_k,\bk$) be the pseudo-rapidity and 
transverse momentum of the parent and the emitted gluon respectively, with $\abar (\eta_p-\eta_k) \ll 1$.
The cross-section for inclusive two-gluon production can be computed as
 \begin{align}
 \label{samey}
 \frac{\dif \sigma_{2g}}{\dif \eta_p\dif^2 \bp \, \dif \eta_k \dif^2 \bk} =
 \frac{1}{(2\pi)^4} \int_{\bx\bbx} \rme^{-\rmi \bp \cdot (\bx - \bbx)} 
 \big\langle H_{\rm prod}(\bk) \hat{S}_{12}(\bx\bbx)  \big|_{\bar{U}=U}
 \big\rangle_Y,
 \end{align}
with the production Hamiltonian \cite{Hentschinski:2005er,Kovner:2006ge,Kovner:2006wr}
 \begin{align}
 \label{hprod}
 H_{\rm prod}(\bk) = \frac{1}{4\pi^3}\,
 \int_{\by\bby}
 \rme^{-\rmi \bk \cdot (\by - \bby)}
 \int_{\bu\bv}
 \mcal{K}^i_{\by\bu} \, \mcal{K}^i_{\bby\bv}
 \big[L^a_{\bu} - U^{\dagger ab}_{\by} R^b_{\bu}\big]
 \big[\bar{L}^a_{\bv} - \bar{U}^{\dagger ac}_{\bby} \bar{R}^c_{\bv}\big]
 \end{align}
and where the ordering of the derivatives is again irrelevant as in the case of the JIMWLK Hamiltonian in \eqn{jimwlk}. Clearly, the functional derivatives denoted with a bar act on Wilson lines in 
the CCA. The four diagrams generated by the action of $H_{\rm prod}$ in 
\eqn{samey} are shown in Fig.~\ref{fig:twogluons}.
Note that after acting with $H_{\rm prod}$ one sets $\bar{U}=U$ and then one performs
the average over $U$ with CGC weight-function $W_Y[U]$ according to \eqn{oave}. 
For instance the $R \bar{R}$ term becomes  
 \begin{align}
 \label{rr}
 \big\langle R^b_{\bu}\, \bar{R}^c_{\bv}\, \hat{S}_{12}(\bx\bbx)\big|_{\bar{U}=U} \big\rangle_Y
 = \int [DU] W_Y[U]\, 
 \frac{1}{N_g}\, 
 \rmTr \big[ \big(R^c_{\bv} U_{\bbx}^\pd\big) 
 \big(R^b_{\bu} U^{\dagger}_{\bx} \big)\big],
 \end{align}
where in the r.h.s.~there is no `bar' anymore, neither on the Wilson lines nor in the functional derivatives. It is a
straightforward exercise to perform the various derivatives, then integrate over $\bu$ and $\bv$, and
thus obtain (up to a factor $\delta(1-x_p-x_k)$ expressing the conservation of the plus component of
the momentum; here, $x_p$ and $x_k$ are the longitudinal momentum fractions taken by the
two final gluons and are such that $x_k\ll 1$ and $x_p\simeq 1$)
 \begin{align}
 \label{twogluon}
 &\frac{\dif \sigma_{2g}}{\dif \eta_p\dif^2 \bp \, \dif \eta_k \dif^2 \bk} =
 \frac{1}{(2\pi)^4}
 \frac{\abar}{\pi} \int_{\bx\bbx\by\bby}
 \rme^{-\rmi \bp \cdot (\bx - \bbx)-\rmi \bk \cdot (\by - \bby)}
 \mcal{K}^i_{\by\bx} \, \mcal{K}^i_{\bby\bbx}\,
 \bigg\langle
 \frac{1}{N_g}\rmTr[U_{\bbx}^\pd U^{\dagger}_{\bx}]
 \nn 
 &\hspace*{1.8cm}-
 \frac{1}{N_c N_g} \big(U^{\dagger}_{\by}+ U^{\dagger}_{\bby}\big)^{ab}
 \rmTr[U^{\dagger}_{\bx} T^b U_{\bbx}^\pd T^a]
 + \frac{1}{N_c N_g}
 \big(U^\pd_{\bby} U^{\dagger}_{\by}\big)^{ab}
 \rmTr[U^{\dagger}_{\bx} T^b T^a U_{\bbx}^\pd]
 \bigg\rangle_Y,
 \end{align}
in agreement with the calculation in \cite{Iancu:2013dta}.

The above manipulations can be easily generalized to the production of several gluons having
similar rapidities. For instance, in order to produce two gluons with transverse momenta $\bk_A$
and $\bk_B$ and such that gluon $A$ is softer than gluon $B$ ($\eta_B > \eta_A$, but such that
$\abar (\eta_p-\eta_A) \ll 1$ and $\abar (\eta_p-\eta_B) \ll 1$), one needs to act twice with the
production Hamiltonian on the generating functional and only then set $\bar{U}=U$, as follows:
$H_{\rm prod}(\bk_A) H_{\rm prod}(\bk_B) \hat{S}_{12}(\bx\bbx)  \big|_{\bar{U}=U}$.
That is, the softer gluon (here, gluon $A$) is produced after the harder one (gluon $B$), and in
particular it can be emitted by the latter: the functional derivatives inside $H_{\rm prod}(\bk_A)$
can also act on the Wilson lines which enter the structure of $H_{\rm prod}(\bk_B)$.

Returning to the cross-section for two gluon production, \eqn{twogluon}, notice that
if we integrate over $\bp$ and over $\eta_p$, that is, we do not measure the parent gluon after the scattering,
the first phase factor in the integrand of \eqn{twogluon} leads to $(2\pi)^2 \delta^{(2)}(\bbx-\bx)$ in \eqn{twogluon} 
and one recovers the well-known result for single gluon production out of gluon source \cite{Kovchegov:1998bi}
 \begin{align}
 \label{single}
 \frac{\dif \sigma_{1g}}{\dif \eta_k \dif^2 \bk} = 
 \frac{1}{(2\pi)^2}\,\frac{\abar}{\pi}
 \int_{\bx\by\bby}
 \rme^{-\rmi \bk \cdot (\by - \bby)}
 \mcal{K}_{\by\bby\bx}\,
 \langle 1 - \hat{S}_{\bx\bby} - \hat{S}_{\by\bx} + \hat{S}_{\by\bby} \rangle_Y.
 \end{align}
 
If on the other hand, one integrates \eqn{twogluon} over $\bk$ and $\eta_k$, that is, one measures
just the parent gluon, then one obtains a part of the radiative corrections to the cross-section in
\eqn{ptbroad} for single inclusive gluon production --- namely, the `real' part associated with the 
emission of a gluon which is not measured. The complete radiative corrections (to leading logarithmic
accuracy at high energy) also include `virtual' processes, i.e.~diagrams where the evolution gluon
is emitted and reabsorbed on the same side of the cut. For the single gluon production in \eqn{ptbroad} 
and also for the production of two gluons with roughly the same rapidity, the quantum evolution 
with increasing energy is controlled by the usual Balitsky-JIMWLK evolution of the Wilson line correlators 
which enter Eqs.~\eqref{ptbroad} or  \eqref{twogluon}. But in the case where the produced gluons are
widely separated in rapidity, we need a generalization of the Balitsky-JIMWLK equations, to be described
in the next section.

\section{Production of gluons at different rapidities}
\label{sect-different}

\begin{figure}
\begin{center}
\includegraphics[width=0.7\textwidth,angle=0]{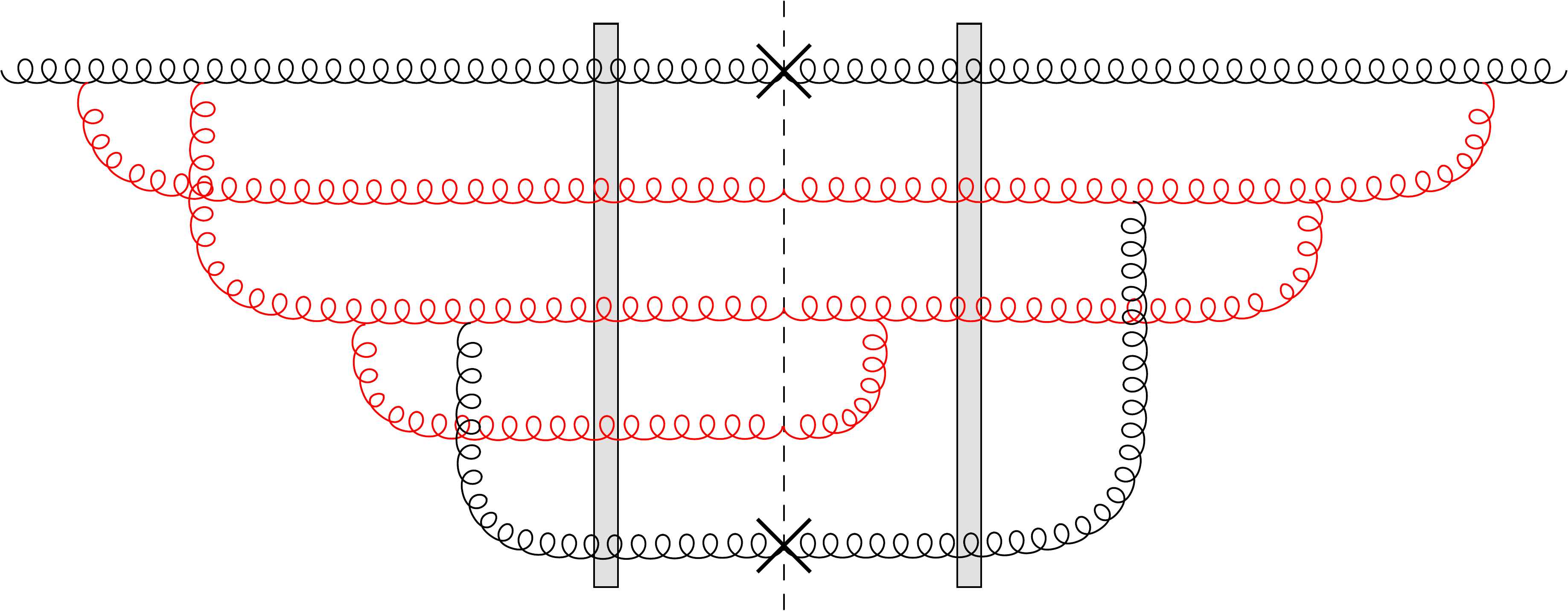}\,
\end{center}
\caption{A typical diagram for the production of two gluons at different rapidities.}
\label{fig:evol_prod}
\end{figure}

We now come to the main problem of interest here,
which is the production of two gluons at widely separated rapidities, 
in which case many more (unresolved) gluons can be emitted between the two measured ones, 
as shown in Fig.~\ref{fig:evol_prod}. 
The first gluon is assumed to be close to the projectile rapidity (that is, its rapidity difference
w.r.t.~the target is equal to $Y$) and thus can be viewed as the ``parent'' gluon. The second, ``emitted'', gluon
has a relative rapidity $Y_A$ w.r.t.~the target,  such that $\abar (Y-Y_A) > 1$. In order to emit this gluon we
need to act with the corresponding production Hamiltonian $H_{\rm prod}^A$ on the WFS of the projectile 
evolved from $Y_A$ up to $Y$. Here, $H_{\rm prod}^A$ is given by an expression similar to \eqn{hprod},
but where the Wilson lines and the functional derivatives refer to the rapidity $Y_A$; we shall
denote the corresponding Wilson lines as $U_A$ (in the DA) and $\bar{U}_A$ (in the CCA).
Accordingly, we need to evolve the generating functional of the projectile gluon from rapidity $Y_A$ 
to rapidity $Y$, with an initial condition at $Y=Y_A$ given by \eqn{ggf} with $U\to U_A$ and $\bar{U}\to \bar{U}_A$.
The answer reads 
 \begin{align}
 \label{gfevol}
 \big\langle \hat{S}_{12}(\bx\bbx) \big\rangle_{Y-Y_A}^A
 = \int [D U D \bar{U}]\, 
 W_{Y-Y_A}[U,\bar{U}|U_A,\bar{U}_A]\,
 \frac{1}{N_g}\,
 \rmTr\big[\bar{U}^\pd_{\bbx} U^{\dagger}_{\bx}\big],
 \end{align}
with the superscript $A$ standing for the functional dependence on $U_A$ and $\bar{U}_A$. The QCD evolution 
in the above is encoded in the {\em conditional} CGC weight-function, which satisfies \cite{Hentschinski:2005er}
 \beq
 \label{hevol}
 \frac{\del}{\del Y}\,
 W_Y[U,\bar{U}|U_A,\bar{U}_A] = 
 H_{\rm evol}\, W_Y[U,\bar{U}|U_A,\bar{U}_A]
 \quad \mathrm{with} \quad
 H_{\rm evol} = H_{11} + H_{22} + 2 H_{\rm 12}
 \eeq
where $H_{11}$ is the JIMWLK Hamiltonian given in \eqn{jimwlk}, $H_{22}$ is obtained by putting a bar in all Wilson lines and derivatives in \eqn{jimwlk}, and $H_{12}$ by putting a bar only in the quantities in the second square bracket factor. This conditional weight-function obeys the initial condition 
 \begin{align}
 W_{Y=0}[U,\bar{U}|U_A,\bar{U}_A] = \delta[U-U_A]\, \delta[\bar{U} - \bar{U}_A].
 \end{align}
Moreover,  at any $Y$  it satisfies the following, important, property
 \begin{align}\label{sym}
 W_Y[U,\bar{U}|U_A,U_A] = \delta[U-\bar{U}]\, W_Y[U|U_A],
 \end{align}
where $W_Y[U|U_A]$ is the conditional weight-function associated to the
usual JIMWLK Hamiltonian --- that is, the solution to \eqn{rg} with initial condition 
$W_{Y=0}[U|U_A]= \delta[U-U_A]$. \eqn{sym} states that the DA and the CCA evolve in 
the same way, so the respective field configurations coincide with each other at any $Y$ provided 
they do so in the initial condition at $Y=0$. This is of course the same property which allows one
to relate cross-sections for particle production like Eqs.~\eqref{twogluon} and \eqref{single} to forward
scattering amplitudes (hence, to total cross-sections).

 \begin{figure}
\begin{center}
\includegraphics[width=0.48\textwidth,angle=0]{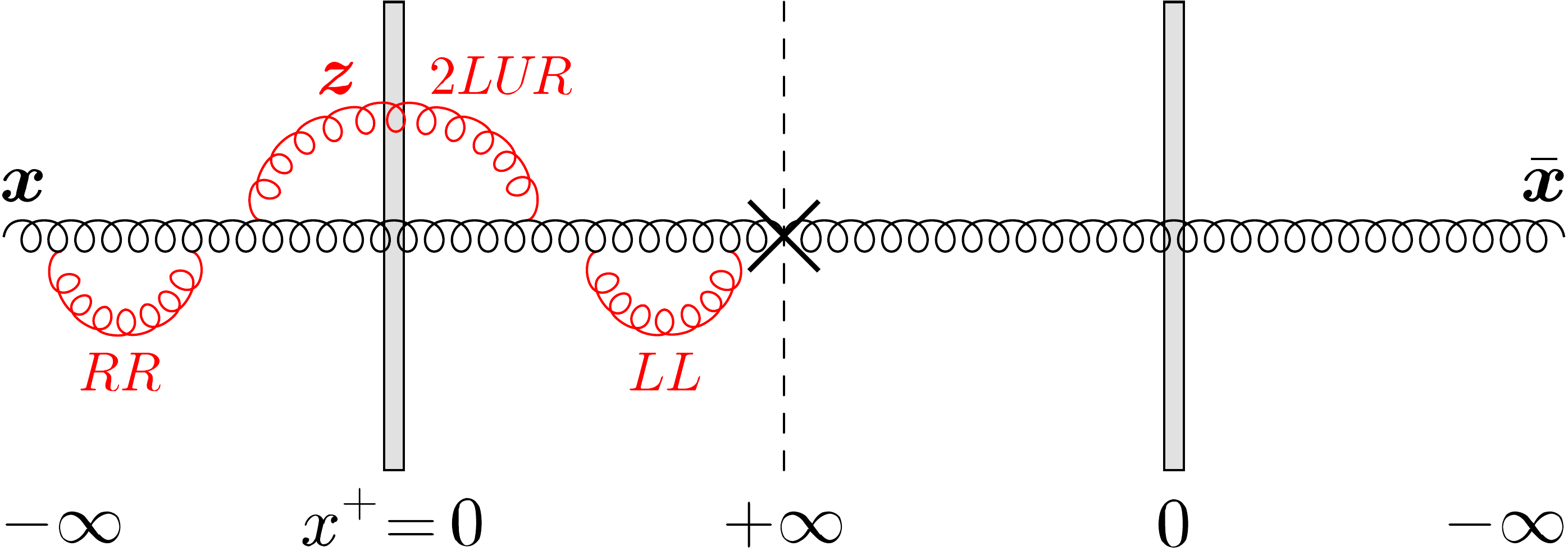}\quad 
\includegraphics[width=0.48\textwidth,angle=0]{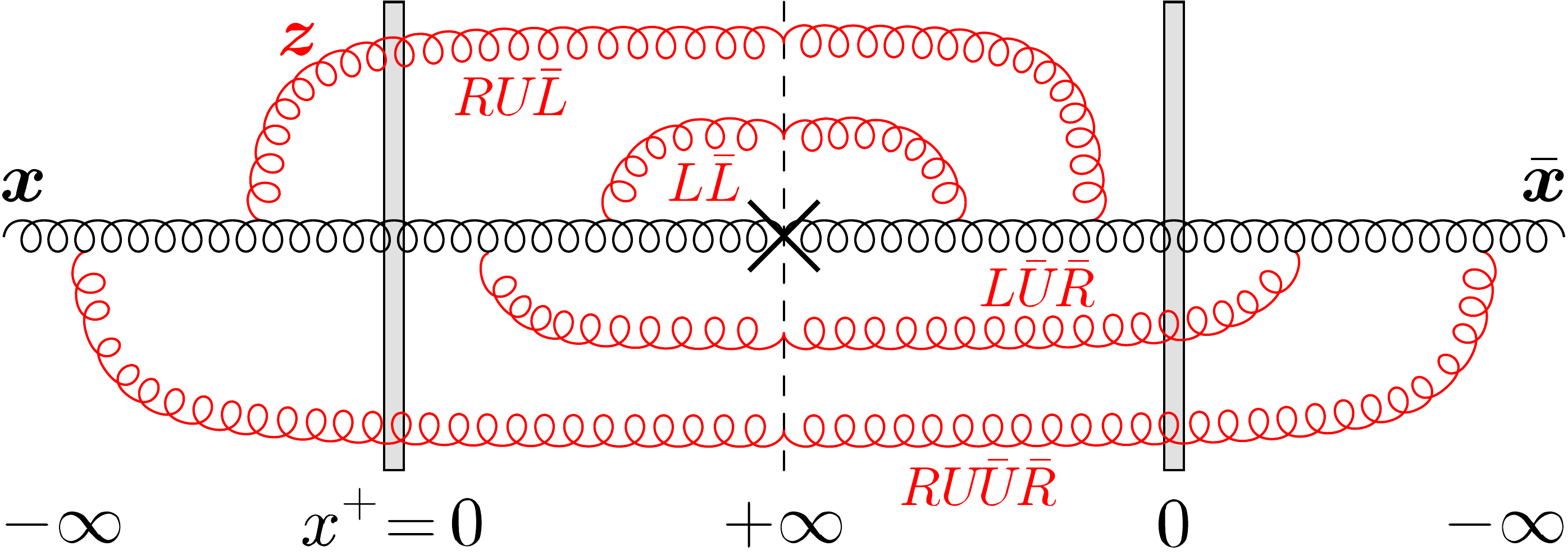}
\end{center}
\caption{The action of $H_{11}$ (left) and $H_{12}$ (right) of $H_{\rm evol}$  on the generating functional
of a projectile gluon. The latter `evolves' by emitting gluons either before, or after, its interaction
with the shockwave representing the target.}
\label{fig:hevol}
\end{figure}

Fig.~\ref{fig:hevol} illustrates the classes of diagrams generated by the action of $H_{\rm evol}$
on the WFS of a bare gluon.
$H_{11}$ and $H_{22}$ produce virtual graphs, whereas $H_{12}$ generates real gluon emissions,
which are not resolved (the emitted gluons have the same transverse coordinate in the DA
and the CCA). It is easier to interpret these diagrams in terms of {\em projectile} evolution, 
i.e.~by assuming that the evolution step is achieved by boosting the projectile gluon, and not
the nuclear target. This will be shortly discussed in more detail.

The corresponding cross-section can be given a formal expression in terms of a path integral:
first we act with $H_{\rm prod}^A\equiv H_{\rm prod}(\bk_A)[U_A,\bar{U}_A]$ on the generating functional in \eqn{gfevol}, then we let $\bar{U}_A = U_A$ and finally we average over $U_A$ with the CGC weight-function 
$W_{Y_A}[U_A]$~:
  \begin{align}
 \label{diffy}
 \frac{\dif \sigma_{2g}}{\dif Y \dif^2 \bp \, \dif Y_A \dif^2 \bk_A} \propto
 \int& [D U_A]\, W_{Y_A}[U_A]
 \int [D\bar{U}_A]\, \delta[\bar{U}_A -U_A]
 \nn
 &H_{\rm prod}^A
 \int[DU D\bar{U}]\, 
 W_{Y-Y_A}[U,\bar{U}|U_A,\bar{U}_A]\,
 \frac{1}{N_g}\,
 \rmTr\big[\bar{U}^\pd_{\bbx} U^{\dagger}_{\bx}\big],
 \end{align}
where the proportionality constant (including the Fourier transform) can be read from \eqn{samey}. 
This equation can be suggestively and more succinctly rewritten as
 \begin{align}
 \label{diffy1}
 \frac{\dif \sigma_{2g}}{\dif Y \dif^2 \bp \, \dif Y_A \dif^2 \bk_A} \propto
\Big\langle H_{\rm prod}^A \big\langle \hat{S}_{12}(\bx\bbx)
  \big\rangle_{Y-Y_A} \big|_{\bar{U}_A=U_A} \Big\rangle_{Y_A}\,,
  \end{align}
where the notation emphasizes the fact that the present calculation involves two types of target averaging: 
one involving the conditional CGC weight-function, which encodes the evolution from 
$Y_A$ up to $Y$, and one using the standard weight-function, for the evolution from $Y_{\rm in}$ 
up to $Y_A$. 

\begin{figure}
\begin{center}
\includegraphics[width=0.47\textwidth,angle=0]{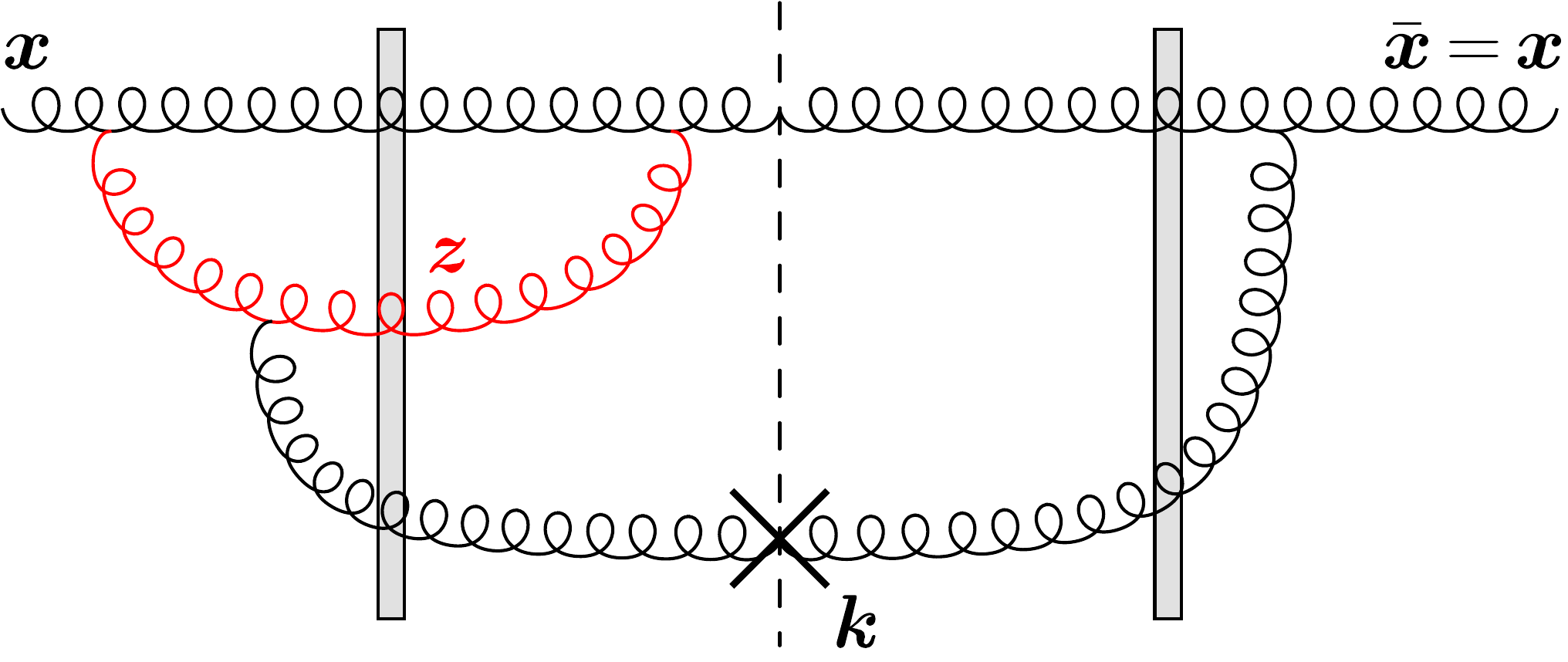}
\qquad
\includegraphics[width=0.47\textwidth,angle=0]{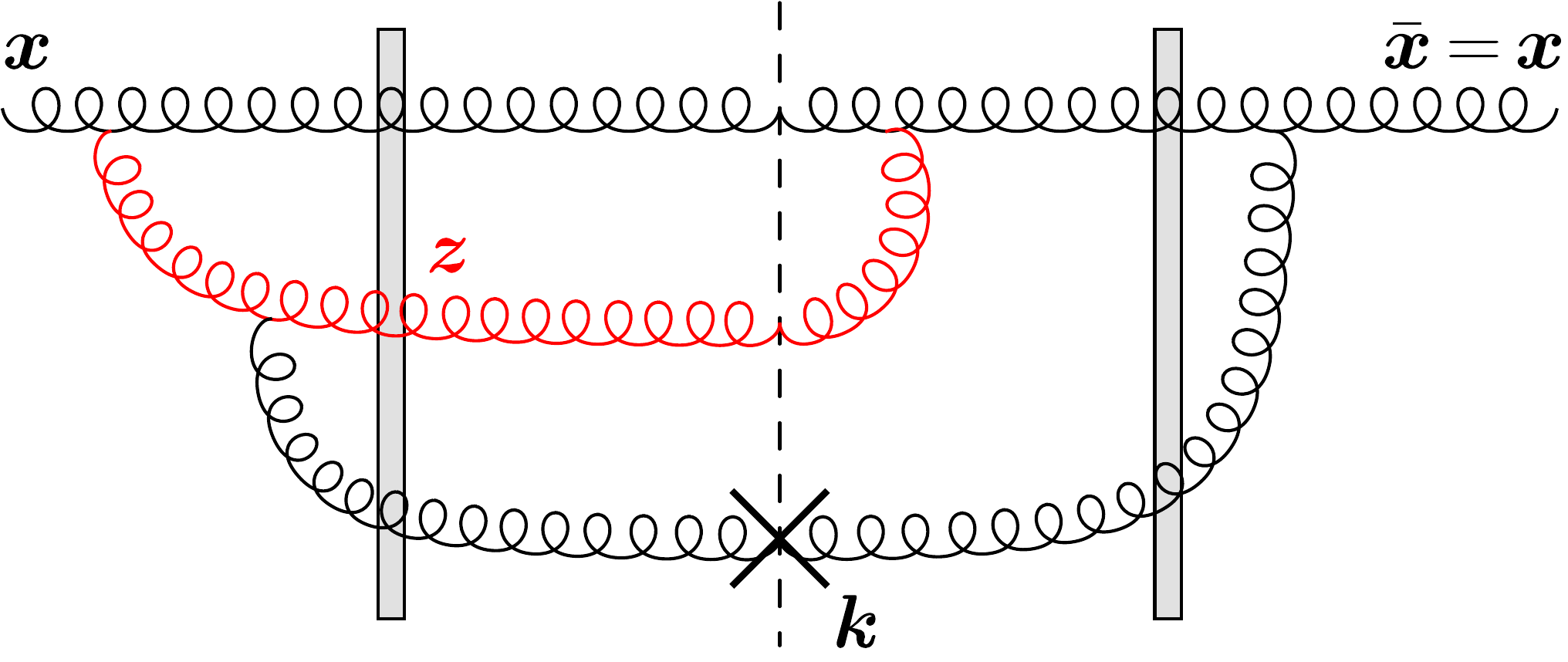}
\end{center}
\caption{Cancellations in the final-state evolution of the ``parent'' gluon when it is not measured. The two diagrams, generated by $RUL$ and $RU\bar{L}$, add to zero since they can be obtained from one another by simply moving a gluon vertex attachment from the DA to the CCA.}
\label{fig:fsi}
\end{figure}

So far, we have privileged the viewpoint of target evolution, as manifest on equations 
like \eqref{gfevol} and \eqref{diffy}. This turns out to be more fruitful for our present purposes,
and we shall indeed stick to it in most of our subsequent developments. But for the
sake of the physical interpretation, it is useful to notice that the evolution 
from $Y_A$ up to $Y$ can alternatively be viewed as the
BFKL evolution of the projectile in the presence of the strong target field
(evolved up to rapidity $Y_A$). This appears to be
more complicated than the usual BFKL evolution of the gluon distribution in the projectile, 
in that it involves both  `initial-state' and `final-state' emissions:  the evolution gluons
can be emitted and/or reabsorbed both before and after the scattering with the shockwave,
as illustrated in Fig.~\ref{fig:hevol}. This introduces a dependence
upon the background field (the Wilson line $U_A$), via processes in which the evolution gluon crosses
the shockwave only once; such processes are generated by terms like $LUR$ or $L U\bar{R}$  
in $H_{\rm evol}$. For the processes where the gluon crosses the shockwave twice, as
generated by $RU\bar{U}^\dagger\bar{R}$, the Wilson lines cancel between the DA and the CCA,
by unitarity. (More precisely, this cancellation occurs after `producing' the
gluon at rapidity $Y_A$, that is, after acting with $H_{\rm prod}^A$ and then letting $\bar{U}_A=U_A$.)
But the usual BFKL evolution is recovered in the case where the parent gluon is not measured,
as recovered by chosing $\bbx=\bx$ in the previous formul\ae. In that case, the 
final-state evolution cancels out between the DA and the CCA, as illustrated in Fig.~\ref{fig:fsi}.
This cancellation occurs already at the level of the generating functional, i.e.~before acting
with $H_{\rm prod}^A$~: for $\bbx=\bx$, the evolution of the generating functional in \eqn{gfevol}
is controlled solely by the terms involving `right' derivatives in $H_{\rm evol}$ --- that is, the 
terms proportional to $RR$, $\bar{R}\bar{R}$, and $R\bar{R}$ ---, which describe the initial-state
evolution of the projectile. This is required by causality: as their name suggest, the `final-state emissions' 
occur after the gluon at rapidity $Y_A$ has been produced and hence they cannot influence its
emission.

The simplest way to render manifest the above physical picture is by
constructing evolution equations for quantities like $\big\langle \hat{S}_{12}(\bx\bbx)
  \big\rangle_{Y-Y_A}$ and $H_{\rm prod}^A \big\langle \hat{S}_{12}(\bx\bbx)
  \big\rangle_{Y-Y_A}$ --- the projectile generating functional and respectively the cross-section for 
producing a gluon at $Y_A$. Such equations can be derived in the same way as the Balitsky equations:
starting with the CGC representation \eqref{gfevol} for the generating functional, one takes a derivative
w.r.t.~$Y$, uses the generalized JIMWLK equation \eqref{hevol}, and then integrates the action 
of $H_{\rm evol}$ by parts, to make the functional derivatives act on $\hat{S}_{12}(\bx\bbx)$.
The evolution equations thus obtained generalize the Balitsky-JIMWLK hierarchy to multi-particle 
production. The first few equations in this new hierarchy will be presented, together with
their physical interpretation, in Appendix \ref{app-gfhier}. (At large $N_c$, similar equations have been previously
constructed in Refs.~\cite{JalilianMarian:2004da,Kovner:2006wr}.)
But these equations appear to be even more involved than the original Balitsky equations.
Although some simplifications occur at large $N_c$ (where only dipoles and quadrupoles remain as the 
relevant degrees of freedom; see  Appendix \ref{app-gfhier}), it appears very difficult to make 
progress with them in practice. 
In what follows, we shall return to the viewpoint of target evolution and propose a Langevin formulation for 
expressions like \eqn{diffy}, that we expect to be better suited for numerical solutions.  

\section{The Langevin description of multi-particle production}
\label{sect-langevin}

To arrive at a Langevin formulation for the two-gluon cross-section in \eqn{diffy}, we start with
\eqn{gfevol} for the evolution of the projectile generating functional over the intermediate rapidity range
$Y-Y_A$. In an analogous way to \eqn{sgdipave}, one can write
 \begin{align}\label{S12Lan}
 \big\langle \hat{S}_{12}(\bx\bbx) \big\rangle_{Y-Y_A}^A 
 = \frac{1}{N_g}\,
 \big \langle \rmTr\big[\bar{U}^\pd_{N_A,\bbx} U^{\dagger}_{N_A,\bx}\big] 
 \big\rangle_{\nu}.
 \end{align}
Here, the rapidity interval is discretized as $Y-Y_A = \epsilon N_A$ and the Wilson lines are built starting at $n=0$, corresponding to $Y_0 \equiv Y_A$, with $U_0 \equiv U_A$ and $\bar{U}_0 \equiv \bar{U}_A$. Specifically,
the Wilson lines in the DA are built according to Eqs.~\eqref{ulang} -- \eqref{noise}, while for the CCA
we similarly write
 \begin{align}
 \label{ublang}
 \bar{U}^{\dagger}_{n,\bx} = 
 \exp[\rmi \epsilon g \bar{\alpha}^{L}_{n,\bx}]\,
 \bar{U}^{\dagger}_{n-1,\bx} \exp[-\rmi \epsilon g \bar{\alpha}^{R}_{n,\bx}],
 \end{align}   
where the left and right matrix fields now read
 \begin{align}
 \label{alphabl}
 &\bar{\alpha}^{L}_{n,\bx} = 
 \frac{1}{\sqrt{4\pi^3}}\int_{\bz} \mcal{K}^i_{\bx\bz} \,\nu^{ia}_{n,\bz}\, T^a,
 \\
 \label{alphabr}
 &\bar{\alpha}^{R}_{n,\bx} = 
  \frac{1}{\sqrt{4\pi^3}}\int_{\bz} \mcal{K}^i_{\bx\bz} \,\nu^{ia}_{n,\bz} 
 \bar{U}^{\dagger ab}_{n-1,\bz}\, T^b.
 \end{align}
Importantly, the noise term $\nu_n$ in these stochastic equations is exactly the same as in
the corresponding equations, \eqref{alphal} -- \eqref{alphar}, for the DA --- in particular, 
it satisfies \eqn{noise}. This is in agreement 
with the fact that, as already discussed, the evolution of the DA and that of the CCA are
strictly correlated with each other: a gluon which is emitted say in the DA can be then
absorbed either in the DA, or in the CCA. It is furthermore consistent with the structure of the 
evolution Hamiltonian in \eqn{hevol}, as it can be explicitly checked. (For instance, one can verify 
that Eqs.~\eqref{ublang} -- \eqref{alphabr}, together with Eqs.~\eqref{ulang} -- \eqref{noise} and \eqn{hevol}, lead to the same evolution equations for the generating 
functionals as obtained in  Appendix \ref{app-gfhier}.)
With reference to Eqs.~\eqref{alphabl} -- \eqref{alphabr}, this means that the `left' matrix field in the CCA 
is the same as that in the DA ($\bar{\alpha}^{L}_{n,\bx} = \alpha^{L}_{n,\bx}$), while the `right' one 
in the CCA ($\bar{\alpha}^{R}_{n,\bx}$) differs from the corresponding
one in the DA (${\alpha}^{R}_{n,\bx}$) only through its dependence on the Wilson line of the previous step. 
Therefore any difference between the `barred' and  `unbarred' 
Wilson lines at the final rapidity $Y$ can be traced
back to a difference between the respective initial conditions ($U_A$ and $ \bar{U}_A$) 
at rapidity $Y_A$.

Although mathematically well defined, the stochastic process that we have just described is not
well suited for numerical simulations, because of the need to keep trace of the functional
dependence upon the initial Wilson lines $U_A$ and $ \bar{U}_A$. 
Yet, as we now explain, one can set up an alternative Langevin process, which is oriented towards
the physical problem at hand --- that is, it depends upon the specific structure of the observable 
(here, the cross-section for two gluon production) --- and which circumvents this problem: in
this new process, the initial Wilson lines are {\em (random) numbers} (more properly, 
numerically-valued color matrices), and not {\em functions}.

To introduce this alternative procedure, let us recall that the evaluation of the cross-section of interest
requires the action of $H_{\rm prod}^A$ on the evolved generating functional in \eqn{S12Lan}.
This in turn involves  the linear combination of 4 terms, such as 
 \begin{align}
 \label{rr2}
 R^a_{A,\bu}\, \bar{R}^b_{A,\bv}\, \big\langle \hat{S}_{12}(\bx\bbx) \big\rangle_{Y-Y_A}^A
 \big|_{\bar{U}_A=U_A}
 = \frac{1}{N_g}\,
 \big\langle 
 \rmTr \big[ \big(R^b_{A,\bv} U_{N_A,\bbx}^\pd\big) 
 \big(R^a_{A,\bu} U^{\dagger}_{N_A,\bx} \big)\big]
 \big\rangle_{\nu},
 \end{align}
where we have added the subscript $A$ on the Lie derivatives to emphasize
that they are acting on the functional dependence of the Wilson lines $U^{\dagger}_{N_A} $
and $\bar{U}_{N_A}$ upon their initial conditions $U_A^{\dagger}$ and respectively $ \bar{U}_A$. 
Note that the difference between  `barred' and  `unbarred' quantities has disappeared
in the r.h.s.~of \eqn{rr2} --- this was only needed to guide the action of
the derivatives. Thus, for the purpose
of evaluating this expression (and the 3 similar ones generated by 
the other pieces, $L^a_{A} \bar{L}^a_{A}$, $R^a_{A} \bar{L}^b_{A}$, and
$L^a_{A} \bar{R}^b_{A}$, of $H_{\rm prod}^A$), 
there is no need to distinguish between the DA and the CCA anymore.
Yet, one still needs to cope with the functional dependence upon the initial conditions,
which is the problem alluded to above. Our proposal to circumvent this problem is to enlarge
the Langevin process, in such a way to include the building blocks of expressions like
\eqn{rr2} --- that is, $R_A U^\dagger$ and $L_A U^\dagger$.

To this aim, one must supplement the Langevin equation \eqref{ulang} for the Wilson lines,
which is {\em local} in the transverse plane, with the following, {\em bilocal},
recurrence formula for $R_A U^\dagger$,
 \begin{align}
 \label{ronu}
 \hspace*{-0.4cm}
 R^a_{A,\bu} U^{\dagger}_{n,\bx} =\,
 &\exp[\rmi \epsilon g \alpha^{L}_{n,\bx}]
 \big(R^a_{A,\bu} U^{\dagger}_{n-1,\bx}\big) \exp[-\rmi \epsilon g \alpha^{R}_{n,\bx}]
 \nn
 &-\frac{\rmi \epsilon g  }{\sqrt{4\pi^3}} \,
 \exp[\rmi \epsilon g \alpha^{L}_{n,\bx}] U^{\dagger}_{n-1,\bx}\int_{\bz} 
  \mcal{K}^i_{\bx\bz}
 \big[U^\pd_{n-1,\bz} \nu^{i}_{n,\bz} U^{\dagger}_{n-1,\bz},
 U^\pd_{n-1,\bz} R^a_{A,\bu} U^{\dagger}_{n-1,\bz} \big],
 \end{align}
with $\nu^{i}_{n,\bz}\equiv  \nu^{i b}_{n,\bz} T^b$ the noise color matrix in the adjoint representation. 
This equation follows from \eqn{ulang} after using the Leibniz rule for differentiation together
with a few manipulations that we shall now explain. The emergence of the first term in the r.h.s.
being pretty obvious, let us concentrate on the second term, which expresses the action
of $R^a_{A}$ on the `right' infinitesimal color rotation. To obtain this term, we have kept only the
linear term in the expansion of $\exp[-\rmi \epsilon g \alpha^{R}_{n,\bx}]$ : indeed, the zeroth order term
(the unity) gives trivially zero under the action of $R^a$, and the same is also true for the quadratic term,
which is effectively independent of $U^{\dagger}_{n-1}$ to the accuracy of interest
(since it can be averaged over the noise; recall the discussion after \eqn{ulang}). After this expansion,
one is led to evaluate (cf. \eqn{alphar})
 \begin{align}
R^a_{A,\bu} \Big(\nu^{ic}_{n,\bz} U^{\dagger cb}_{n-1,\bz}\, T^b\Big)&\,=\,
R^a_{A,\bu} \Big(U^\pd_{n-1,\bz} \nu^{i}_{n,\bz} U^{\dagger}_{n-1,\bz}\Big)\nn
&\,=\,\big(R^a_{A,\bu} U^\pd_{n-1,\bz}\big) \nu^{i}_{n,\bz} U^{\dagger}_{n-1,\bz}
+ U^\pd_{n-1,\bz} \nu^{i}_{n,\bz}\big(R^a_{A,\bu} U^{\dagger}_{n-1,\bz}\big)
\nn&\,=\,\big[U^\pd_{n-1,\bz} \nu^{i}_{n,\bz} U^{\dagger}_{n-1,\bz},
 U^\pd_{n-1,\bz} R^a_{A,\bu} U^{\dagger}_{n-1,\bz} \big]
 \end{align}
where we have also used the identities $U^{\dagger cb} T^b = U T^c U^{\dagger}$ and  
$R^a U = - U (R^a U^{\dagger}) U$ (with the latter arising from $R^a(UU^{\dagger}) = 0$).

In writing  \eqn{ronu}, we have used the same discretization conventions as in relation with \eqn{S12Lan}.
That is, we have written $Y-Y_A = \epsilon N_A$ and used $n$ with $n=1,2, \dots N_A$
to denote a generic intermediate step. The initial condition, represented by $n=0$, 
corresponds to $Y_0 \equiv Y_A$ and $U_0 \equiv U_A$, and reads
  \begin{align}
 \label{ronuzero}
 R^a_{A,\bu} U^{\dagger}_{A,\bx}
 = \rmi g \delta_{\bu\bx} U^{\dagger}_{A,\bx} T^a\,,\qquad
 L^a_{A,\bu} U^{\dagger}_{A,\bx} =
 \rmi g \delta_{\bu\bx} T^a U^{\dagger}_{A,\bx}\,. \end{align}
The quantity $L^a_{A,\bu} U^{\dagger}_{n,\bx}$ obeys an equation identical to \eqn{ronu}, 
but with a different initial condition, as also indicated above. However, there is no need
to separately consider the associated equation, because of the relation between 
`left' and `right' Lie derivatives mentioned after \eqn{comm}: once that the quantity
$R^a_{A,\bu} U^{\dagger}_{N_A,\bx}$ is obtained by iterating \eqn{ronu}, one can
immediately deduce $L^a_{A,\bu} U^{\dagger}_{N_A,\bx} = U^{\dagger ab}_{A,\bu}
\big(R^b_{A,\bu} U^{\dagger}_{N_A,\bx}\big)$. Also, one does not need  to separately  calculate 
$R^a_{A,\bu} U_{n,\bx}$ : it is an easy exercise to prove by induction that it can be obtained by 
taking the hermitian conjugate of $R^a_{A,\bu} U^{\dagger}_{n,\bx}$. To conclude, we just need
to follow two independent Langevin processes: \eqn{ulang} for the Wilson line $U^{\dagger}_{n,\bx} $
and \eqn{ronu} for its `right' Lie derivative $R^a_{A,\bu} U^{\dagger}_{n,\bx}$.
 
\eqn{ronu} introduces a numerical complication as compared to \eqn{ulang}: unlike \eqn{ulang},
which is {\em local} in the transverse coordinates, \eqn{ronu} is {\em bi-local}. Actually, the corresponding
initial condition is still local, because of the presence of the factor $\delta_{\bu\bx}$ in \eqn{ronuzero}, but
already its first iteration acquires a bi-local structure, as one can easily check. In spite of 
this potential complication, which is purely numerical,
the procedure that we have just outlined has a decisive advantage over \eqn{S12Lan}, namely
it allows for a fully numerical implementation, at least in principle. Indeed,
the initial condition $U_A$ for \eqn{ronu} is not a {\em generic} color matrix anymore, 
but the {\em numerical} matrix that has been generated in the previous steps 
of the Langevin process, from $Y=Y_{\rm in}$ (where $U=U_{\rm in}$ is randomly 
selected according to the Gaussian MV distribution), up to $Y=Y_A$ (where $U=U_A$).
Of course, during the first part of this evolution, from $Y_{\rm in}$ up to $Y_A$, there is no 
analog of  \eqn{ronu} : this part involves just the usual Langevin process,
\eqn{ulang}, for the Wilson lines.

Note also an alternative form for the Langevin equation \eqref{ronu}, which is perhaps
more convenient for numerics, in that it involves the infinitesimal right rotation 
$\exp[\rmi \epsilon g \alpha^{R}_{n,\bx}]$ alone\footnote{We thank T.~Lappi for suggesting us to look for such an alternative rewritting.}:
 \begin{align}
 \label{uronu}
 U^\pd_{n,\bx} R^a_{A,\bu} U^{\dagger}_{n,\bx} =\,
 &\exp[\rmi \epsilon g \alpha^{R}_{n,\bx}]
 \big(U^\pd_{n-1,\bx} R^a_{A,\bu} U^{\dagger}_{n-1,\bx}\big)
 \exp[-\rmi \epsilon g \alpha^{R}_{n,\bx}]
 \nn
 &-\frac{\rmi \epsilon g  }{\sqrt{4\pi^3}} 
 \exp[\rmi \epsilon g \alpha^{R}_{n,\bx}]
 \int_{\bz} 
 \mcal{K}^i_{\bx\bz}
 \big[U^\pd_{n-1,\bz} \nu^{i}_{n,\bz} U^{\dagger}_{n-1,\bz},
 U^\pd_{n-1,\bz} R^a_{A,\bu} U^{\dagger}_{n-1,\bz} \big].
 \end{align}
Let us observe here that \eqn{uronu} should be consistent with the fact that $U^\pd_{n,\bx} R^a_{A,\bu} U^{\dagger}_{n,\bx}$ must be a member of the Lie algebra. This is manifest for the first term on the r.h.s.~of \eqn{uronu}, but it is not so obvious for the second term due to the presence of a \emph{single} infinitesimal right rotation. However, one should keep in mind that the equation is valid only up to order $\epsilon$ and thus, recalling that the noise scales like $1/\sqrt{\epsilon}$, one can expand $\exp[\rmi \epsilon g \alpha^{R}_{n,\bx}]$ to first order. Then the second term of \eqn{uronu} becomes
 \begin{align}
 \label{uronuexpand}
 &-\frac{\rmi \epsilon g  }{\sqrt{4\pi^3}}
 \int_{\bz} 
 \mcal{K}^i_{\bx\bz}
 \big[U^\pd_{n-1,\bz} \nu^{i}_{n,\bz} U^{\dagger}_{n-1,\bz},
 U^\pd_{n-1,\bz} R^a_{A,\bu} U^{\dagger}_{n-1,\bz} \big]
 \nn
 &+
 \frac{\epsilon g^2}{4\pi^3}
 \int_{\bz} 
 \mcal{K}_{\bx\bx\bz}\,
 U^\pd_{n-1,\bz} T^b U^{\dagger}_{n-1,\bz}
 \big[U^\pd_{n-1,\bz} T^b U^{\dagger}_{n-1,\bz},
 U^\pd_{n-1,\bz} R^a_{A,\bu} U^{\dagger}_{n-1,\bz} \big],
 \end{align}
where we have been allowed to take the average for the contribution quadratic in the noise. It is now clear that both terms in \eqn{uronuexpand} are members of the Lie algebra.
 
To summarize, by iterating the recurrence formula \eqn{ulang} over the whole rapidity interval
between $Y_{\rm in}$  (where we insert the initial condition of the MV type) up to $Y$,
and the new formula \eqref{ronu} (or \eqref{uronu}) over the upper rapidity range between $Y_A$ and $Y$, 
one can construct the r.h.s.~of \eqn{rr2} via a fully numerical procedure. The only potential difficulty that
we can foresee is a numerical complication associated with the bi-local structure of \eqn{ronu}
in transverse coordinates.
At this point it is should be clear that the method can be extended to the production of more than two gluons, but it becomes more and more tedious (since increasingly non-local) with increasing the number of measured gluons. 
As an illustration, we shall outline the corresponding procedure for three gluon production in 
Appendix \ref{app-three}.

\section*{Acknowledgments}
We would like to thank Al Mueller for stimulating discussions and Tuomas Lappi 
for helpful comments and for reading the manuscript. 
All figures were made with Jaxodraw \cite{Binosi:2003yf}.
This work is supported by the Agence Nationale de la Recherche project \# 11-BS04-015-01.

\appendix
\section{Evolution equations for the projectile generating functionals}
\label{app-gfhier} 

As discussed at the end of Sect.~\ref{sect-different}, one method to study the evolution of the 
generating functionals with $Y$ is to construct the respective evolution equations, which 
generalize the Balitsky hierarchy to multi-particle production. These equations are
obtained by acting on the generating functionals with the evolution Hamiltonian $H_{\rm evol}$ in 
\eqn{hevol}. Here, we are interested only in illustrating the procedure, and for this purpose
we shall consider the simplest generating functional, that of a physical quark. 
This is represented by a mathematical dipole, similar to that in \eqn{ggf},  
but where the Wilson lines are now in the fundamental representation and will be denoted by $V$.
Differentiating the quark analog of \eqn{gfevol} w.r.t.~$Y$, using \eqn{hevol}, 
and integrating by parts, we find
 \begin{align}
 \label{gffevol}
 \frac{\del \big\langle \hat{S}^F_{12}(\bx\bbx) \big\rangle_{Y}}{\del Y}
 =  \frac{1}{N_c} \big\langle 
 H_{\rm evol}\rmTr\big[\bar{V}^\pd_{\bbx} 
 V^{\dagger}_{\bx}\big] \big\rangle_Y,
 \end{align}   
with $F$ standing for the fundamental representation. A straightforward exercise, similar to the derivation of the Balitsky
hierarchy from the JIMWLK equation, leads to the following equation (after also using 
the Fierz identities to rearrange the terms)
 \begin{align}
 \label{gffevol2}
 \frac{\del \big\langle \hat{S}^F_{12}(\bx\bbx) \big\rangle_{Y}}{\del Y}
 = \frac{\abar}{2\pi}
 \int_{\bz}
 \Big[& 
 (\mcal{K}_{\bx\bx\bz} - \mcal{K}_{\bx\bbx\bz})
 \big\langle \hat{S}^F_{11}(\bx\bz) \hat{S}^F_{12}(\bz\bbx) 
 -\hat{S}^F_{12}(\bx\bbx) \big\rangle_{Y}
 \nn
 & + (\mcal{K}_{\bbx\bbx\bz} - \mcal{K}_{\bx\bbx\bz})
 \big\langle \hat{S}^F_{12}(\bx\bz) \hat{S}^F_{22}(\bz\bbx) 
 - \hat{S}^F_{12}(\bx\bbx) \big\rangle_{Y}
 \nn
 & + \mcal{K}_{\bx\bbx\bz}
 \big\langle 
 \hat{S}^F_{12}(\bz\bz) \hat{Q}^F_{1221}(\bx\bbx\bz\bz)
 - \hat{S}^F_{12}(\bx\bbx) \big\rangle_{Y} \Big].
 \end{align}
In $\hat{S}^F_{11}$ both Wilson lines live in the DA, while in $\hat{S}^F_{22}$ they both live in the CCA, that is,
\beq \hat{S}^F_{11}(\bx\bz) \equiv  \frac{1}{N_c} \,
\rmTr\big[{V}^\pd_{\bz} 
 V^{\dagger}_{\bx}\big] \,,\qquad  \hat{S}^F_{22}(\bz\bbx) \equiv  \frac{1}{N_c} \,
\rmTr\big[\bar{V}^\pd_{\bbx} \bar{V}^\dagger_{\bz}\big]\,.\eeq
The mathematical quadrupole $\hat{Q}^F_{1221}$ stands for the generating functional of a physical 
quark-antiquark dipole, and reads 
 \begin{align}
 \label{quad}
 \hat{Q}^F_{1221}(\bx\bbx\bbz\bz) = 
 \frac{1}{N_c} \rmTr
 \big[V^\pd_{\bz} 
 \bar{V}^{\dagger}_{\bbz}\bar{V}^\pd_{\bbx} 
 V^{\dagger}_{\bx}\big],
 \end{align}
where  $\bx$ and $\bz$ are the coordinates of the quark and the antiquark in the DA, while
$\bbx$ and $\bbz$ similarly refer to the CCA.

Below, we shall be interested in emitting a gluon at rapidity $Y_A$ out of a quark with rapidity $Y>Y_A$. 
For that purpose, \eqn{gffevol2} has to be integrated from $Y_A$ up to $Y$,
with a functional initial condition at $Y_A$~:
$\big\langle \hat{S}^F_{12}(\bx\bbx) \big\rangle_{Y_A} = (1/N_c)
\rmTr\big[\bar{V}^\pd_{A,\bbx} V^{\dagger}_{A,\bx}\big]$. 

Clearly, \eqn{gffevol2} is not a closed equation,
rather it involves new generating functionals in the r.h.s., for which we need to construct the
respective equations. This complication should not be a surprise, given our experience
with the Balitsky hierarchy. But \eqn{gffevol2} looks considerably more involved than
the corresponding Balitsky equation (that for the dipole $S$--matrix), both because of the
complex structure of its r.h.s.~and because of the functional initial conditions.
Like in the Balitsky hierarchy, important simplifications occur at large $N_c$. But before 
we discuss them, let us notice some special cases of \eqn{gffevol2} :

(a) If one sets $\bar{V}_A = V_A$ in the initial conditions, that is, if one identifies the Wilson lines in 
the DA and the CCA at the initial rapidity, then this property remains true at any $Y>Y_A$, cf.
\eqn{sym}, and then there is no difference anymore 
between $\hat{S}^F_{11}$, $\hat{S}^F_{22}$, and $\hat{S}^F_{12}$. In that case,
one can easily see that the last term in \eqn{gffevol2} vanishes, while the first two combine to give the `BK equation'
(more properly, the first equation in the Balitsky hierarchy). In the present context, this equation describes
the evolution with $Y$ of the cross-section for single inclusive quark production (cf.  \eqn{ptbroad}).
But of course, after identifying $\bar{V}=V$, one loses\footnote{Alternatively, one can say
that one loses the `quantum fluctuations', by which we here mean the gluons which can be emitted
or absorbed by the projectile and which are encoded in the (functional) difference between
$\bar{V}$ and $V$. In the Keldysh-Schwinger formalism, this would refer to the difference between
the quantum fluctuations on the two branches of the closed time contour.}
the very meaning of a  `generating functional' : one cannot use $\hat{S}^F_{12}$ to
emit gluons anymore.

(b) The case where the parent quark is not resolved is obtained by setting $\bbx = \bx$, and then the first two terms in \eqn{gffevol2} individually vanish. This is the expected cancellation of the final state evolution between the DA and the CCA (cf.  Fig.~\ref{fig:fsi}). Correspondingly, the surviving term in the third line is 
associated with initial state evolution alone. Indeed, one can verify that this term could 
directly be obtained by keeping only the `right' derivatives in $H_{\rm evol}$, that is, 
the three terms which involve $RR$, $\bar{R}\bar{R}$, and $R\bar{R}$.

Since the `quadrupole' generating functional in \eqn{quad} 
(the WFS of a physical dipole) enters the evolution equation \eqref{gffevol2}  for the `dipolar' one 
(the WFS of a physical quark), 
it is useful to also have a look at the corresponding evolution equation. After some algebra,
this is found as
 \begin{align}
 \label{qgfevol}
 \frac{\del \big\langle \hat{Q}^F_{1221}(\bx\bbx\bbz\bz) \big\rangle_{Y}}{\del Y}
 = \frac{\abar}{4\pi}
 \int_{\by} 
 &(\mcal{M}_{\bx\bz\by} + \mcal{M}_{\bx\bbx\by} - \mcal{M}_{\bbx\bz\by})
 \big\langle \hat{S}^F_{11}(\bx\by) 
 \hat{Q}^F_{1221}(\by\bbx\bbz\bz) \big \rangle_Y
 \nn
 +&
 (\mcal{M}_{\bx\bz\by} + \mcal{M}_{\bz\bbz\by} - \mcal{M}_{\bx\bbz\by})
 \big\langle \hat{S}^F_{11}(\by\bz) 
 \hat{Q}^F_{1221}(\bx\bbx\bbz\by) \big \rangle_Y
 \nn*[0.2cm]
 +&
 (\mcal{M}_{\bbx\bbz\by} + \mcal{M}_{\bx\bbx\by} - \mcal{M}_{\bx\bbz\by})
 \big\langle \hat{S}^F_{22}(\by\bbx) 
 \hat{Q}^F_{1221}(\bx\by\bbz\bz) \big \rangle_Y
 \nn*[0.2cm]
 +&
 (\mcal{M}_{\bbx\bbz\by} + \mcal{M}_{\bz\bbz\by} - \mcal{M}_{\bbx\bz\by})
 \big\langle \hat{S}^F_{22}(\bbz\by) 
 \hat{Q}^F_{1221}(\bx\bbx\by\bz) \big \rangle_Y
 \nn*[0.2cm]
 -&
 (\mcal{M}_{\bx\bz\by} + \mcal{M}_{\bbx\bbz\by} + 
 \mcal{M}_{\bx\bbx\by} + \mcal{M}_{\bz\bbz\by})
 \big\langle \hat{Q}^F_{1221}(\bx\bbx\bbz\bz) \big \rangle_Y
 \nn*[0.2cm]
 - &
 ( \mcal{M}_{\bx\bz\by} + \mcal{M}_{\bbx\bbz\by}
 -\mcal{M}_{\bbx\bz\by} - \mcal{M}_{\bx\bbz\by})
 \big\langle \hat{S}^F_{11}(\bx\bz) \hat{S}^F_{22}(\bbz\bbx)  \big \rangle_Y
 \nn*[0.2cm]
 &\hspace{-2.2cm} -
 (\mcal{M}_{\bx\bbx\by} + \mcal{M}_{\bz\bbz\by}
 -\mcal{M}_{\bbx\bz\by} - \mcal{M}_{\bx\bbz\by})
 \big\langle \hat{Q}^F_{1221}(\bx\bbx\by\by) \hat{Q}^F_{1221}(\by\by\bbz\bz)  \big \rangle_Y\,,
 \end{align}
where $\mcal{M}$ is the `dipole kernel',
 \beq
 \mcal{M}_{\bu\bv\bz}\,\equiv\,\mcal{K}_{\bu\bu\bz}+\mcal{K}_{\bv\bv\bz}
 -2\mcal{K}_{\bu\bv\bz}\,=\,
  \frac{(\bm{u}-\bm{v})^2}
 {(\bm{u}-\bm{z})^2(\bm{z}-\bm{v})^2}\,,\eeq
which vanishes when $\bu=\bv$.
The initial condition at $Y=Y_A$ is given by \eqn{quad} with $V\to V_A$ and $\bar{V}\to \bar{V}_A$.
As before one can recognize a couple of special cases in \eqn{qgfevol}~: 

(a) In the limit of where one identifies 
$\bar{V} = V$ (`no quantum fluctuations'), one recovers the standard Balitsky equation for the evolution
of the $S$--matrix of a {\em physical} quadrupole \cite{JalilianMarian:2004da,Kovner:2006ge}. In the present context, this equation could refer e.g.~to the cross-section for producing a 
quark-antiquark pair in deep inelastic scattering off the nucleus \cite{Dominguez:2011wm}.

 (b) If the parent quark and antiquark are not resolved, one sets $\bbx = \bx$ and $\bbz = \bz$, and then
 \eqn{qgfevol} reduces to its very last term, which describes the evolution of the WFS of a $q\bar q$ color
 dipole via initial state emissions alone. In particular at large $N_c$ one can factorize
 $\langle \hat{Q}^F\hat{Q}^F \rangle \simeq \langle \hat{Q}^F \rangle \langle \hat{Q}^F \rangle$ (see below),
 and then one recovers, as expected, the evolution equation for the generating functional of an `onium'
 (the system of dipoles generated via the BFKL evolution of an original dipole,
 in the limit where $N_c\gg 1$) \cite{Mueller:1993rr}.
 
We are now in a position to describe the simplifications which occur at large $N_c$. First, expectation
values of products of color traces can be factorized into products of averages of the individual traces. 
This property is well known to hold for the solutions to the Balitsky-JIMWLK equations provided it is already
satisfied by the initial conditions at $Y=Y_{\rm in}$ (as indeed happens within the MV model at large $N_c$). The
respective argument can be taken over to the evolution Hamiltonian $H_{\rm evol}$ in \eqn{hevol} 
and it implies a similar factorization for the generating functionals;
e.g.~$\big\langle \hat{S}^F_{11}(\bx\bz) \hat{S}^F_{12}(\bz\bbx)  \big\rangle_{Y}\simeq
\big\langle \hat{S}^F_{11}(\bx\bz)\big\rangle_{Y} \big\langle \hat{S}^F_{12}(\bz\bbx)  \big\rangle_{Y}$.
Moreover, still at large $N_c$, the quantities $\big\langle \hat{S}^F_{11}\big\rangle_{Y}$ and
$\big\langle \hat{S}^F_{22}\big\rangle_{Y}$ lose their meaning as generating functionals, because the
two functional derivatives inside $H_{\rm prod}$ must act on Wilson lines from a same color trace,
to generate the maximal power of $N_c$. Hence, inside these quantities one can set $\bar{V} = V$
already before acting with $H_{\rm prod}$, and then they reduce to the standard dipole $S$--matrix
(the solution to BK equation).
Similar simplifications apply to  \eqn{qgfevol} for the quadrupole. 

We conclude that, in the large $N_c$ 
limit, the correspondingly  simplified versions of Eqs.~\eqref{gffevol2} and  \eqref{qgfevol} (with $\bbz=\bz$)
form a $2 \times 2$ system for $\langle \hat{S}_{12} \rangle_Y$ and $\langle \hat{Q}_{1221} \rangle_Y$. 
In that sense, multi-particle production at large $N_c$ involves only dipoles and quadrupoles, as already
noticed in the literature \cite{JalilianMarian:2004da,Kovner:2006wr,Dominguez:2012ad}. In fact,
the large $N_c$ versions of Eqs.~\eqref{gffevol2} and \eqref{qgfevol} are equivalent to the respective 
equations constructed in Refs.~\cite{JalilianMarian:2004da,Kovner:2006wr} directly at large $N_c$.
The coefficients in these equations also involve the dipole $S$-matrix $\big\langle \hat{S}^F \big \rangle_{Y}$,
which here describes multiple scattering for the (unresolved)
gluons associated with final-state evolution.

But even at large $N_c$, the equations above have the drawback that they must be solved with
functional initial conditions, which is not very convenient in practice. It is therefore interesting
to notice that {\em ordinary} evolution equations ---
i.e.~equations for the production cross-sections which describe the evolution from $Y_A$
up to $Y$ (at large $N_c$) and admit numerical initial conditions at $Y_A$ ---
 can be deduced by acting with $H_{\rm prod}^A$ on the above equations. 
 We recall that the derivatives inside $H_{\rm prod}^A$ act on the functional dependence
of the WFS upon the Wilson lines $V_A$ and $\bar{V}_A$, as introduced via the initial conditions at $Y_A$.
Also, after acting with $H_{\rm prod}^A$, one has to set $V_A=\bar{V}_A$ (which ensures that 
$\bar{V} = V$ at any $Y>Y_A$) and then average over $V_A$ with
the CGC weight-function $W_{Y_A}[V_A]$.

To be more precise, let us consider the evolution of the cross-section for quark-gluon production.  Starting with the
large $N_c$ version of \eqn{gffevol} and performing the manipulations explained above, one
finds  \begin{align}
 \label{nkevol}
 \frac{\del \big\langle N_{\bk}(\bx\bbx) 
 \big\rangle_{Y}}{\del Y}
 = \frac{\abar}{2\pi}
 \int_{\bz}
 \Big\{& 
 (\mcal{K}_{\bx\bx\bz} - \mcal{K}_{\bx\bbx\bz})
 \big[\big\langle \hat{S}^F(\bx\bz) \big\rangle_{Y}
 \big\langle N_{\bk}(\bz\bbx) \big\rangle_Y 
 -\big\langle N_{\bk}(\bx\bbx) \big\rangle_{Y}\big]
 \nn
  +& (\mcal{K}_{\bbx\bbx\bz} - \mcal{K}_{\bx\bbx\bz})
 \big[\big\langle \hat{S}^F(\bz\bbx) \big\rangle_{Y}
 \big\langle N_{\bk}(\bx\bz) \big\rangle_Y 
 -\big\langle N_{\bk}(\bx\bbx) \big\rangle_{Y}\big]
 \nn
 &\hspace{-1.5cm} + \mcal{K}_{\bx\bbx\bz}
 \big[\big\langle \hat{S}^F(\bx\bbx) \big\rangle_{Y}
 \big\langle N_{\bk}(\bz\bz) \big\rangle_Y 
 -\big\langle N_{\bk}(\bx\bbx) \big \rangle_Y
 + \big\langle N_{\bk}^{(2)}(\bx\bbx\bz\bz)\big\rangle_{Y}\big]
 \Big\},
 \end{align}
where $\big\langle \hat{S}^F \big\rangle_{Y}$ is the solution to the BK equation
and
 \begin{align}
  \big\langle N_{\bk}(\bx\bbx) 
 \big\rangle_{Y}  &\equiv  \Big\langle H_{\rm prod}^A(\bk) \big\langle \hat{S}^F_{12}(\bx\bbx)
  \big\rangle_{Y-Y_A} \big|_{\bar{V}_A=V_A} \Big\rangle_{Y_A}\,,\nn
  \big\langle N_{\bk}^{(2)}(\bx\bbx\bbz\bz)\big\rangle_{Y} &\equiv   \Big\langle
  H_{\rm prod}^A(\bk) 
  \big\langle\hat{Q}^F_{1221}(\bx\bbx\bbz\bz) \big\rangle_{Y-Y_A} \big|_{\bar{V}_A=V_A}
  \Big\rangle_{Y_A}\,,\end{align}
For more clarity, we have indicated through our notations that the generating functionals
have to be evolved from $Y_A$ up to $Y$. Also, we have used the `double--averaging'
notation introduced in \eqn{diffy1}. Up to an overall factor and a Fourier transform, 
$\big\langle N_{\bk}(\bx\bbx)  \big\rangle_{Y}$ is the cross-section for quark-gluon production,
with the quark at rapidity $Y$ and the gluon at rapidity $Y_A < Y$ (cf.~\eqn{samey}).
Similarly, $ \big\langle N_{\bk}^{(2)}(\bx\bbx\bbz\bz)\big\rangle_{Y}$ is proportional to the
cross-section for producing a gluon with momentum $\bk$ at rapidity $Y_A$ out of a $q\bar q$
dipole at rapidity $Y$ and such that the quark and the antiquark can be measured as well.
To get closed equations, one also needs the evolution equation obeyed
by the last quantity, at least for the case where the antiquark is not measured ($\bbz=\bz$). 
This follows from \eqn{qgfevol} and reads
 \begin{align}
 \label{nk2evol}
 \frac{\del \big\langle N_{\bk}^{(2)}(\bx\bbx\bz\bz) \big\rangle_{Y}}{\del Y}
 = \frac{\abar}{4\pi}
 \int_{\bw} 
 &(\mcal{M}_{\bx\bz\bw} + 
 \mcal{M}_{\bx\bbx\bw} - 
 \mcal{M}_{\bbx\bz\bw})
 \big\langle \hat{S}^F(\bx\bw) \big \rangle_Y
 \big \langle N_{\bk}^{(2)}(\bw\bbx\bz\bz) \big \rangle_Y
 \nn
 +&
 (\mcal{M}_{\bbx\bz\bw} + 
 \mcal{M}_{\bx\bbx\bw} - 
 \mcal{M}_{\bx\bz\bw})
 \big\langle \hat{S}^F(\bw\bbx) \big \rangle_Y
 \big\langle N_{\bk}^{(2)}(\bx\bw\bz\bz) \big \rangle_Y
 \nn*[0.1cm]
 -&
 (\mcal{M}_{\bx\bz\bw} + 
 \mcal{M}_{\bbx\bz\bw} + 
 \mcal{M}_{\bx\bbx\bw})
 \big\langle N_{\bk}^{(2)}(\bx\bbx\bz\bz)\big \rangle_Y
 \nn*[0.1cm]
 + &
 (\mcal{M}_{\bbx\bz\bw} + 
 \mcal{M}_{\bx\bz\bw} - 
 \mcal{M}_{\bx\bbx\bw})
 \big\langle N_{\bk}^{(2)}(\bx\bbx\bw\bw)\big \rangle_Y
 \nn*[0.1cm]
 + &
 (\mcal{M}_{\bbx\bz\bw} + 
 \mcal{M}_{\bx\bz\bw} - 
 \mcal{M}_{\bx\bbx\bw})
 \big\langle \hat{S}^F(\bx\bbx) \big \rangle_Y
 \big\langle N_{\bk}^{(2)}(\bw\bw\bz\bz)  \big \rangle_Y.
 \end{align}
When one also sets $\bbx=\bx$, i.e.~when neither the quark nor the antiquark are being measured, 
the first two terms in the r.h.s of this equation individually vanish, while the surviving terms reproduce
the BFKL equation for $N_{\bk}^{(2)}(\bx\bx\bz\bz)$, as expected: the cross-section for producing
a gluon out of the onium evolves with the energy in the same way as the dipole (or gluon)
distribution of the onium, since the gluon can be emitted
by any of the dipoles composing the onium. Note that the `color transparency' property 
$\big\langle \hat{S}^F(\bx\bx) \big \rangle_Y=1$ has been important too for this argument 
(it has been used to simplify the last term in \eqn{nk2evol}). In the present
context, this property expresses the cancellation of the `final-state interactions' of the unmeasured  quark  
--- i.e.~its interactions with the nuclear target which occur after emitting the gluon  ---
between the DA and the CCA.

Eqs.~\eqref{nkevol}  and \eqref{nk2evol} must be integrated from $Y_A$ up
to $Y$, with initial conditions given by the respective cross-sections (for quark-gluon production and
respectively dipole-gluon production). These equations are linear w.r.t.~the quantities that
one needs to solve for, i.e.~the cross-sections $\big\langle N_{\bk}(\bx\bbx)  \big\rangle_{Y}$ and  
$\big\langle N_{\bk}^{(2)}(\bx\bbx\bz\bz)\big\rangle_{Y}$, but in general they include
multiple scattering effects via the dipole $S$-matrix $\big\langle \hat{S}^F \big \rangle_{Y}$.
Such effects are associated with the final-state evolution, or the final-state interactions, of
the parent quark and disappear when $\bbx=\bx$, i.e.~when this quark is not measured\footnote{There
is of course another limit in which the effects of multiple scattering can be neglected, including
for the case where the parent partons {\em are} measured, i.e.~for multi-particle production:
this is the limit where the target itself is dilute, as e.g.~in proton-proton collisions at not too high energies. 
The corresponding limit of Eqs.~\eqref{nkevol}  and \eqref{nk2evol}
is obtained by replacing $\big\langle \hat{S}^F \big \rangle_{Y}\to 1$ for
all the dipole $S$-matrices which appear in these equations. This amounts to neglecting the target 
rescattering for both the projectile partons and the evolution gluons. The only interactions
with the target which survive in this limit are those of the produced gluon at the lowest
rapidity $Y_A$. They enter the solutions to the simplified equations via the respective initial 
conditions at $Y=Y_A$, themselves computed in the single scattering approximation. 
After this replacement, the equations become fully linear and describe the
evolution of the BFKL Green's function (for quark and dipole impact
factors respectively).}.
In that limit, Eqs.~\eqref{nkevol}  and \eqref{nk2evol} describe the BFKL evolution of the 
unintegrated gluon distribution in the quark and respectively dipole projectile, at large $N_c$.

To summarize, the general strategy in this case would be as follows: \texttt{(i)} First construct approximate solutions
to the usual Balitsky-JIMWLK equations for the dipole and the quadrupole $S$--matrices, 
$\big\langle \hat{S}^F(\bx\bbx) \big \rangle_{Y}$ and $\big\langle \hat{Q}^F(\bx\bbx\bz\bbz) \big \rangle_{Y}$.
For instance, one can solve the BK equation for $\big\langle \hat{S}^F(\bx\bbx) \big \rangle_{Y}$ and then
use the Gaussian approximation to the JIMWLK evolution to compute the quadrupole in terms of the
dipole.  \texttt{(ii)} Then, use the aforementioned (approximate) solutions in order to compute 
the cross-sections $\big\langle N_{\bk}(\bx\bbx)  \big\rangle_{Y_A}$ and  
$\big\langle N_{\bk}^{(2)}(\bx\bbx\bbz\bz)\big\rangle_{Y_A}$ at rapidity $Y_A$. The respective
expressions are well known: the cross-section
for quark-gluon production looks very similar to \eqn{rr} and can be found in \cite{Marquet:2007vb},
whereas that for dipole-gluon production is presented in \cite{JalilianMarian:2004da}.
\texttt{(iii)}  The cross-sections obtained in the previous step are now used as initial conditions for 
Eqs.~\eqref{nkevol}  and \eqref{nk2evol}, to be integrated from $Y_A$ up to $Y$. Namely, one
needs to first solve  \eqn{nk2evol} for $\langle N_{\bk}^{(2)} \rangle_Y$, then plug the result
into \eqn{nkevol}, and solve the latter for $\langle N_{\bk} \rangle_Y$. \texttt{(iv)} 
Finally one needs to take the Fourier transform as in \eqn{samey}, to ascribe a transverse momentum
$\bp$ to the measured quark.

This being said, this procedure appears as prohibitively cumbersome, including for numerical simulations,
and we believe that the Langevin procedure described in the main text should be more tractable in practice.

\section{Three gluon production}
\label{app-three}

Here we shall discuss the extension of the Langevin procedure to the production of three gluons at ($Y_A,\bk_A$), ($Y_B,\bk_B$) and ($Y,\bp$) with $Y>Y_B>Y_A$. The analog of \eqn{diffy} is
 \begin{align}
 \label{3g}
 \hspace*{-0.3cm}
 \frac{\dif \sigma_{3g}}{\dif Y \dif^2 \bp\, 
 \dif Y_A \dif^2 \bk_A \, \dif Y_B \dif^2 \bk_B} \propto
 \int& [D U_A]\, W_{Y_A}[U_A]
 \int [D\bar{U}_A]\, \delta[\bar{U}_A -U_A]
 \nn
 & H_{\rm prod}^A \int[DU_B D\bar{U}_B]\, 
 W_{Y_B-Y_A}[U_B,\bar{U}_B|U_A,\bar{U}_A]
 \nn
 &H_{\rm prod}^B
 \int[DU D\bar{U}]\, 
 W_{Y-Y_B}[U,\bar{U}|U_B,\bar{U}_B]\,
 \frac{1}{N_g}\,
 \rmTr\big[\bar{U}^\pd_{\bbx} U^{\dagger}_{\bx}\big].
 \end{align}
As in the corresponding discussion in Sect.~\ref{sect-langevin}, this equation is easier to read
by following the evolution backwards in rapidity, from $Y$ down to $Y_{\rm in}$.
For generic Wilson lines $U_B$ and $\bar{U}_B$ at rapidity $Y_B$, one needs to first compute
the action of $H_{\rm prod}^B$ on the generating functional of the projectile gluon evolved
from $Y_B$ up to $Y$~:
 \begin{align}
 \label{hpbons}
 H_{\rm prod}^B\,
 \frac{1}{N_g}\,
 \rmTr
 \big\langle \big[\bar{U}^\pd_{N_B,\bbx} U^{\dagger}_{N_B,\bx}\big]
 \big\rangle_{Y-Y_B}^B.
 \end{align} 
As usual, we consider a discretization of the relevant rapidity interval
according to $Y-Y_B = \epsilon N_B$. The result of this step is itself a functional
of $U_B$ and $\bar{U}_B$. Next, the Wilson lines $U_B$ and $\bar{U}_B$ 
must in turn be built from generic Wilson lines $U_A$ and $\bar{U}_A$, via iterations 
which cover the rapidity interval $Y_B - Y_A = \epsilon N_A$~; as a result, 
\eqn{hpbons} becomes a functional of $U_A$ and $\bar{U}_A$. 
Then we can act with $H_{\rm prod}^A$ and subsequently set $\bar{U}_A = U_A$,
with $U_A$ constructed by evolving the initial condition (say, as given by the MV model)
from $Y_{\rm in}$ to $Y_A$, via the standard JIMWLK evolution. Note that by imposing
$\bar{U}_A = U_A$, we automatically enforce $\bar{U}_B = U_B$, and eventually
$\bar{U}_{N_B}= U_{N_B}$, because of the property \eqref{sym} of the conditional CGC weight-function.

But although conceptually clear, the above procedure --- which would involve only
the standard local Langevin \eqn{ulang}, but with functional initial conditions
at $Y_B$ and $Y_A$ --- is not convenient in practice.
To achieve a fully numerical implementation for the three gluon production, one should rather proceed
{\em upwards} in rapidity and construct in each step all the building blocks (Wilson lines and their
Lie derivatives) which enter the cross-section in \eqn{3g}. In view of the procedure we followed in Sect.~\ref{sect-langevin} we can rewrite \eqn{3g} as
 \begin{align}
 \label{3gf}
 \frac{\dif \sigma_{3g}}{\dif Y \dif^2 \bp\, 
 \dif Y_A \dif^2 \bk_A \, \dif Y_B \dif^2 \bk_B} \propto
 H_{\rm prod}^A \,H_{\rm prod}^B \,
 \frac{1}{N_g} \langle \rmTr [\bar{U}^\pd_{\bbx} U^{\dagger}_{\bx}]\rangle_{\nu}\,.
 \end{align}

At this point let us first recall that `left' Lie derivatives can be written in terms of `right' ones (cf.~the discussions after Eqs.~\eqref{comm} and \eqref{ronuzero}). This means that $H_{\rm prod}$, so long as the square brackets in \eqn{hprod} are considered, can be rewritten as
 \begin{align}
 \label{hprod2}
 H_{\rm prod} \propto
 \big[U^\dagger_{\bu} - U^\dagger_{\by} \big]^{ab}
 \big[\bar{U}^\dagger_{\bv} - \bar{U}^\dagger_{\bby} \big]^{ac}
 R^b_{\bu} \bar{R}^c_{\bv}.
 \end{align}
As in the case of the two-gluon production, it is enough to study the (double) action of $H_{\rm prod}$ on the Wilson line in the DA. Denoting by $\by_{A}$ and $\by_B$ the transverse coordinates of the produced gluons in the DA at $Y_A$ and $Y_B$ respectively, one needs to calculate quantities of the form
 \begin{align}
 \label{hphponu}
 \hspace*{-0.6cm}
 U^{\dagger ab}_{A,\by_A}
 R_{A,\bu}^b
 \big[ U^{\dagger cd}_{B,\by_B} 
 R_{B,\bw}^d U_{\bx}^\dagger \big] = 
 U^{\dagger ab}_{A,\by_A}
 U^{\dagger cd}_{B,\by_B}
 R_{A,\bu}^b R_{B,\bw}^d U_{\bx}^\dagger
 +
 U^{\dagger ab}_{A,\by_A} 
 \big( R_{A,\bu}^b U^{\dagger}_{B,\by_B} \big)^{cd}
 R_{B,\bw}^d U_{\bx}^\dagger,
 \end{align}
where we recall that $R_A$ and $R_B$ are the Lie derivatives w.r.t.~the Wilson lines $U_A$ and $U_B$. All the structures appearing in \eqn{hphponu} can be constructed via a Langevin procedure. Specifically, at $Y=Y_{\rm in}$, one
starts with $U^{\dagger}_{{\rm in},\bx}$, which is randomly selected according to the MV model, 
and uses the Langevin equation \eqref{ulang} to build $U_{A}^{\dagger}$ at $Y=Y_A$. 
Then, using the procedure described in Sect.~\ref{sect-langevin}, one evolves 
from $Y_A$ to $Y_B$, to construct  $U^{\dagger}_{B}$ and  $R_{A} U^{\dagger}_{B}$. Finally one evolves from $Y_B$ up to $Y$ to build $R_{B} U^{\dagger}$ and $R_{A} R_{B}U^{\dagger}$ (during which process, one also needs to follow $U_n^\dagger$ and $R_A U_n^\dagger$).  All these quantities can thus be numerically computed, but the respective Langevin process is more involved than in the case of  two-gluon production: during the uppermost rapidity interval from $Y_B$ to $Y$, the structure $R^b_{A,\bu} R^d_{B,\bw} U^{\dagger}_{n,\bx}$ is tri-local in the transverse plane.

This method considerably simplifies in the case where two of the three 
produced gluons have similar rapidities, say $Y\simeq Y_B$ in \eqn{3g}. Indeed, this avoids
the most complicated step in the above procedure, that of the evolution between $Y_B$ and $Y$.
The corresponding calculation is similar to that of two-gluon production discussed in Sect.~\ref{sect-langevin},
except for the fact that the projectile emitting the soft gluon at rapidity $Y_A$ is not a bare gluon anymore,
but rather a system of two gluons, one of which (the gluon $B$) has been emitted by the other one
(the `parent' gluon). Hence, the generating functional describing the WFS of the emitter is 
not just $\hat{S}_{12}(\bx\bbx)$ anymore, but rather $H_{\rm prod}(\bk_B) \hat{S}_{12}(\bx\bbx)$, 
which is explicitly shown\footnote{More precisely, \eqn{twogluon} is written in the eikonal approximation,
which assumes that gluon $B$ is significantly softer than the parent gluon. When this condition 
is not satisfied, one can replace \eqn{twogluon} by the more general formula in \cite{Iancu:2013dta},
which is only slightly more complicated.} in \eqn{twogluon}. 
The cross-section for producing three gluons with rapidities $Y\simeq Y_B\gg Y_A$ can therefore be 
computed by replacing $\hat{S}_{12}(\bx\bbx)\to H_{\rm prod}(\bk_B) \hat{S}_{12}(\bx\bbx)$ in the procedure
outlined in Sect.~\ref{sect-langevin}, that is, in equations like \eqref{S12Lan} and \eqref{rr2}.


\providecommand{\href}[2]{#2}\begingroup\raggedright\endgroup

\end{document}